%% file: main.tex
\documentclass[conference]{IEEEtran}
\IEEEoverridecommandlockouts

\usepackage{cite}
\usepackage{amsmath,amssymb,amsfonts}
\usepackage{algorithmic}
\usepackage{graphicx}
\usepackage{textcomp}
\usepackage{xcolor}
\usepackage{multirow} 
\usepackage{url}

\usepackage{xcolor}
\usepackage{graphicx}
\usepackage[hidelinks]{hyperref}

\newcommand{\orcidicon}[1]{%
  \href{https://orcid.org/#1}{%
    \raisebox{-0.15ex}{\includegraphics[height=1.6ex]{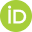}}%
  }%
}

\usepackage{subcaption}
\def\BibTeX{{\rm B\kern-.05em{\sc i\kern-.025em b}\kern-.08em
    T\kern-.1667em\lower.7ex\hbox{E}\kern-.125emX}}
\begin{document}

\title{Human-Flow Digital Twin for Predicting the Effects of Mobility Introduction on Visitor Circulation\thanks{This paper has been accepted for presentation at the 27th IEEE International Conference on Mobile Data Management(MDM2026).}
}

\author{%
  \IEEEauthorblockN{%
    Chiharu Shima\protect\orcidicon{0009-0004-1234-2523}\IEEEauthorrefmark{1},
    Haruki Yonekura\protect\orcidicon{0000-0001-8184-3883}\IEEEauthorrefmark{2}\IEEEauthorrefmark{3},
    Fukuharu Tanaka\protect\orcidicon{0000-0002-8324-9455}\IEEEauthorrefmark{3}\IEEEauthorrefmark{2},
    Tatsuya Amano\protect\orcidicon{0000-0002-8011-247X}\IEEEauthorrefmark{2}\IEEEauthorrefmark{3},
    Hirozumi Yamaguchi\protect\orcidicon{0000-0003-2273-4876}\IEEEauthorrefmark{2}\IEEEauthorrefmark{3}
  }
  \IEEEauthorblockA{%
    \IEEEauthorrefmark{1}bitA Inc., Shinagawa-ku, Tokyo, Japan\\
    \IEEEauthorrefmark{2}The University of Osaka, Suita, Osaka, Japan\\
    \IEEEauthorrefmark{3}RIKEN Center for Computational Science, Kobe, Hyogo, Japan\\
    \{c-shima, h-yonekura, f-tanaka, t-amano, h-yamagu\}@ist.osaka-u.ac.jp
  }
}

\maketitle

\input{00_abstract}


\input{01_introduction_alt}

\input{02_relatedworks}

\input{03_proposed_method_rewrite}

\input{04_experiment_alter}

\input{06_conclusion}

\bibliographystyle{IEEEtran}
\bibliography{ref}

\end{document}

%% file: 00_abstract.tex
\begin{abstract}
We propose a framework for predicting the effects of mobility introduction measures using a human-flow digital twin. 
This digital twin incorporates a multi-agent simulator that can represent how visitors choose destinations depending on factors such as their current location and the attractiveness of spots. 
We extract data on how visitors selected destinations with respect to measured pre-intervention human-flow data, inter-spot distances, spot attractiveness, and travel volumes, and use these data to train each agent’s decision model of this simulator.
The trained decision model is a function that takes a visitor's current state and surrounding environmental information as input and outputs which spot the visitor will move toward next. 
By expressing mobility introduction measures as changes to inter-point distances or to spot attractiveness, the framework can reproduce human flows with mobility introduction in the multi-agent simulator and thereby quantify effects such as changes in visitor counts and circulation.
We evaluated the proposed method using human-flow data measured with and without introducing mobility within Wakayama Castle Park in Japan. 
When reproducing flows with mobility introduction using a multi-layer perceptron decision model, the cosine similarity of the spatial population distribution exceeded 0.7, confirming that the approach can replicate the flow changes caused by the mobility introduction.
\color{black}
\end{abstract}

%% file: 01_introduction_alt.tex
\section{Introduction}

Tourism promotion is widely recognized as an effective means of regional revitalization. 
Local visitor destinations often possess valuable cultural and historical assets, such as heritage buildings, museums, and scenic sites, that can attract visitors. 
However, limited connectivity and the relative difficulty of moving between attractions frequently impede visitor circulation, reducing opportunities for visitors to explore multiple sites and undermining the overall attractiveness and economic potential of these areas.
To mitigate this problem, various new mobility options have been proposed and piloted to facilitate intra-site movement. 
Personal mobility devices, such as e-scooters and e-bikes, and green slow-mobility services, such as low-speed electric shuttle carts and golf-cart-like vehicles, are among the approaches expected to enhance visitor circulation by reducing travel friction and making inter-site movement more convenient \cite{BADIA2023811, yang2021tourists}.

When planning mobility introductions at visitor sites, decision-makers must determine operational parameters such as routing, the spatial placement of boarding and alighting points, service frequency, and fleet size that will most effectively increase visitor circulation. 
Conducting field trials for every candidate design is, however, prohibitively costly and time-consuming. 
Consequently, there is a practical need for predictive tools that can evaluate alternative deployment strategies via simulation before physical implementation, thereby guiding planners toward effective and cost-efficient solutions.
Although prior work has advanced destination-choice modeling and microscopic pedestrian simulation\cite{kaziyeva2023large, horl2021integrating}, few studies explicitly combine learned destination-choice models with region-scale agent-based simulators to evaluate the impacts of introducing new mobility options.

\begin{figure}[t]
    \centering
    \includegraphics[width=\linewidth]{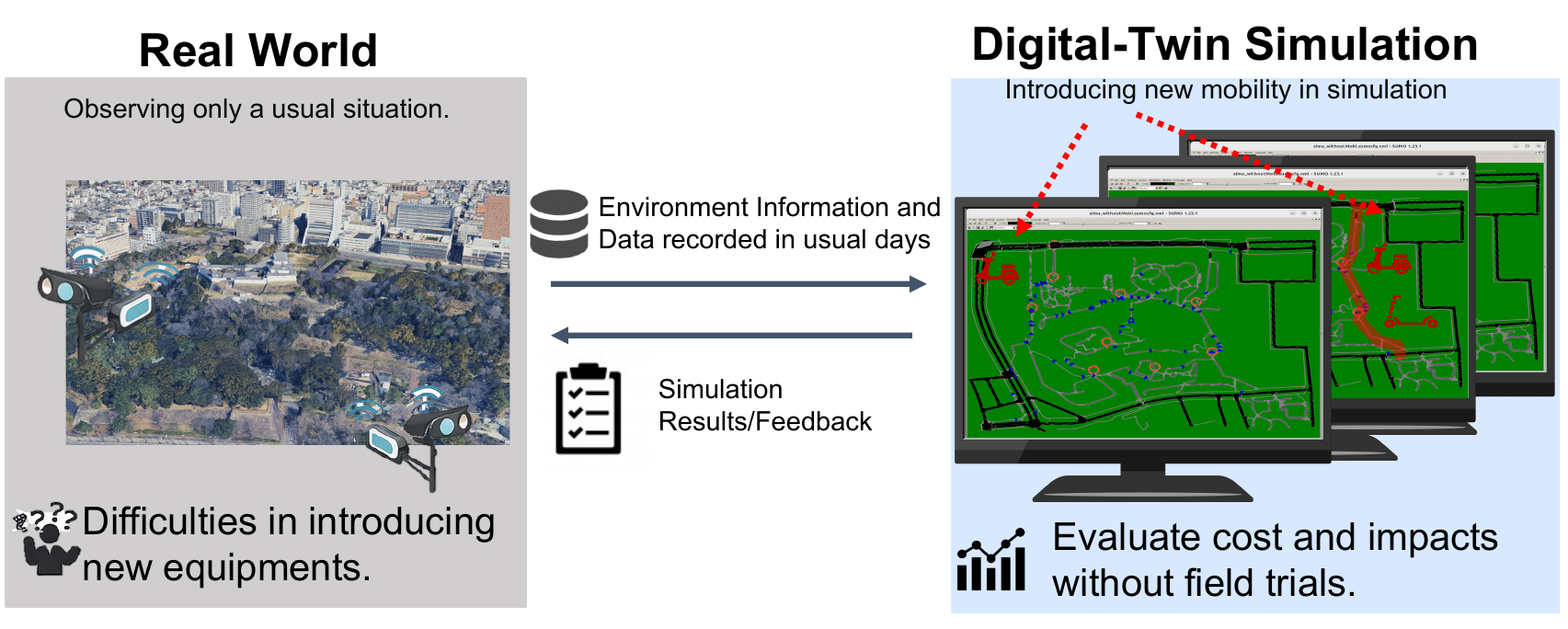}
    \caption{Predicting the Effect of Mobility Introduction via Digital Twin Simulation.}
    \label{fig:graphical_abstract}
\end{figure}

In practice, introducing low-speed mobility services into visitor sites is not merely a matter of adding vehicles. 
Historic parks and cultural districts often contain narrow walkable spaces and mixed-use circulation, so new mobility modes can intensify competition for limited urban space and create site-specific operational and safety challenges. 
Prior work on micro-mobility integration has emphasized that municipalities are often unprepared for such interactions and that data-driven evaluation is important for devising acceptable regulations and countermeasures \cite{dozza2022data}. 
These concerns are particularly salient in tourism settings, where boarding and alighting locations, route design, and pedestrian congestion must be coordinated without undermining visitor comfort or accessibility.

From a planning perspective, the problem is inherently counterfactual: the same site cannot be observed under multiple deployment designs, and reliable intervention-specific data are rarely available before implementation. 
Recent studies have begun to explore counterfactual prediction for tourism planning under privacy-sensitive observations \cite{mishra2024balancing} and to assess the spatial effects of green mobility in urban parks from limited pre-deployment information \cite{naeem2026questionnaire}. 
More broadly, digital twins are attracting increasing attention in transport planning, but their effective application to complex planning processes remains challenging, especially when behavioral adaptation and scenario-based intervention assessment must be integrated within a usable planning workflow \cite{nag2025exploring}. 
These considerations motivate a framework that learns visitor responses from normal-state observations and evaluates mobility introduction as a controlled environmental modification rather than as a separately observed intervention dataset.
In the present implementation, offline trajectory reconstruction and simulation initialization rely on normal-state wide-area location-point sequences that are privacy-safe and anonymized/pseudonymized, but still retain tracking IDs and timestamps, together with site-level spot counts and environmental information. 
Following Uegaki et al.~\cite{ueg2023sim}, these location-point sequences are used to estimate inter-PoI movement priors and relative Origin-Destination (OD) composition for offline initialization, while intervention-specific retraining is avoided at the online prediction stage.

In this paper, we propose a platform that predicts and visualizes the effects of \emph{mobility introduction} (i.e., the deployment of mobility services, such as e-scooters and low-speed electric shuttle carts on designated routes within a visitor site, enabling visitors to move between points of interest) on visitor circulation by means of a human-flow digital twin shown in Fig.~\ref{fig:graphical_abstract}. 
The core of the platform is an agent-based simulation (ABS) that explicitly models the link between individual decision-making and environmental factors. 
Each visitor is represented as an autonomous agent whose destination-choice behavior is governed by a learned decision model. 
Given an agent’s internal state (current location, previously visited points of interest (PoIs), and cumulative walking distance) and environmental inputs (pairwise distances between PoIs, PoI attractiveness, and time of day), the model computes a probability distribution over the agent’s candidate next destinations.
We treat mobility introduction as modifications to the environmental factors that feed into the decision models; by modeling mobility indirectly through such environmental modifications, our approach obviates the need for large, intervention-specific mobility datasets.
By applying the learned decision models within the ABS under such modified environmental conditions, the platform can simulate how the mobility introduction—on selected routes and with specified service parameters—affects visitors' route choices and, in aggregate, the spatio-temporal distribution of visitors.
For the simulation platform, we integrated Simulation of Urban MObility (SUMO) \cite{Dan2002sumo}, JuPedSim \cite{arm2015juped}, and Traffic Control Interface (TraCI) API \cite{axe2008traCI} to capture both high-level decision making and fine-grained pedestrian dynamics. 

We validated the proposed framework using human-flow data collected within Wakayama Castle Park in Japan. 
The park area was partitioned into eight areas, and we evaluated the platform’s ability to reproduce the temporal evolution of per-area population distributions. 
The results indicate that the Multi-Layer Perceptron (MLP) based decision model can reproduce the temporal population distribution under mobility introduction with cosine similarity exceeding 0.7. 
In addition, the MLP model reproduced the mobility-induced change in population distribution with a cosine similarity of 0.664.


%% file: 02_relatedworks.tex
\section{Related Works}
\label{sec:related_work}

\subsection{Destination-Choice Modeling}
Research on destination-choice modeling seeks to explain and predict how individuals select destinations by relating observable environmental factors and latent psychological drivers to choice behavior.
Robin et al. \cite{ROBIN200936} propose discrete-choice formulations that estimate decision probabilities from observational data to model rational destination-selection processes. 
Gao et al. \cite{GAO2023103369} construct a tourism-related knowledge graph and introduce a Knowledge-Graph-aware Disentangled Auto-Encoder (KGDAE) to model tourist choices based on geographic and functional factors. 
Dai et al. \cite{DAI2022104484} define latent motivational constructs such as “travel inspiration” and use these psychological factors to model traveler decision processes. 
McCabe et al. \cite{Mcc2015tour} develop a general model grounded in dual-system theory that emphasizes the role of psychological context and the interplay of personal and contextual factors in shaping choice strategies. 
King et al. \cite{king2022virtual} conduct virtual experiments to investigate whether destination-choice behavior depends on building layout and planned or imposed destination schedules. 
Jin et al. \cite{jin2024exploring} apply machine-learning models and use SHapley Additive exPlanations (SHAP) to analyze and interpret pedestrian route-choice mechanisms.
While these works advance understanding of microscopic choice mechanisms and provide reliable building blocks for demand modeling, they are typically developed and evaluated in settings focused on modeling choice behavior. 
In particular, the literature rarely demonstrates end-to-end propagation of learned destination-choice models into physics-aware, region-scale pedestrian simulations for the purpose of evaluating mobility-mode introductions.

\subsection{Pedestrian Microsimulation}
Microscopic pedestrian simulation research focuses on detailed local movement rules and interaction models that reproduce short-term crowd dynamics and emergent collective behaviors.
Helbing et al. \cite{helbing1991mathematical} 
develop the social force family of models, which represent pedestrian motion through attractive and repulsive interactions and explain emergent patterns such as lane formation and bottleneck oscillations. 
Treuille et al. \cite{treuille2006continuum} propose a continuum-based, real-time crowd simulation method that couples global navigation with local collision avoidance via dynamic potential fields. 
Armel et al. \cite{arm2015juped} present JuPedSim, an open-source framework for simulating and analyzing microscopic pedestrian dynamics. 
Zhang et al. \cite{zhang2021speed} introduce a speed-based pedestrian model that integrates walking comfort and directional preferences into a collision-free framework. 
Dang et al. \cite{dang2024hypedsim} propose an adaptive scheme that switches local motion models according to nearby density to better reproduce individual trajectories.
Such microsimulators are widely used for safety and operational analysis; however, they frequently assume exogenous routing or demand inputs and are not always designed to incorporate empirically-learned, context-sensitive destination-choice behavior directly.

\subsection{Agent-Based Simulation (ABS)}
ABS research aims to bridge individual decision processes and population-level movement by embedding explicit choice models within simulated agents.
Huang et al. \cite{10024374} extend floor-field cellular automata to model decision making under risk and uncertainty in evacuation scenarios. 
Horni et al. \cite{horni2011high} augment MATSim with a high-resolution destination-choice formulation that embeds random-utility components to enable realistic location choice in large-scale agent simulations.
Wörle et al. \cite{WORLE2021202} model intermodal travel behavior and its influence on destination and mode choice within an agent-based travel-demand framework. 
Kaziyeva et al. \cite{kaziyeva2023large} develop a region-scale agent-based simulation that incorporates walkability considerations to better reflect heterogeneous behavior of tourists and residents. 
Yang et al. \cite{yang2024co} propose a co-simulation system that couples MATSim with external fleet and mode simulators to propagate destination-based demand through multimodal simulations. 
Li et al. \cite{li2024agent} describe procedures to generate and validate large-scale agent-based transport models against observed mode shares and travel-time distributions.
Nevertheless, many ABS implementations focus either on macroscopic demand propagation or on transport-mode interactions, and they are not routinely configured to evaluate the introduction of novel, low-speed personal mobility services in a way that jointly preserves micro-level physical interactions and macro-level destination dynamics.

\subsection{Contribution of This Study}
This paper addresses two practical gaps in the related works: the relatively limited attention to modelling the mobility introduction and their impact on visitor circulation, and the lack of a unified, implementable framework that integrates learned destination-choice models with physics-aware microsimulation and end-to-end visualization for site-scale intervention assessment. 
The main contributions of this paper are summarized as follows.
\begin{itemize}
\item We present a human-flow digital twin that embeds learned destination-choice models within an ABS, enabling the reproduction and the visualization of circulation with mobility introduction.
\item We model mobility introduction as changes in environmental features and inject them into learned decision models to predict individual choices. This reduces the need for intervention-specific data, although offline initialization still requires normal-state anonymized/pseudonymized wide-area location sequences with tracking IDs and timestamps, along with spot counts, to estimate movement priors.
\item We deliver a practical simulation platform that couples SUMO, destination-choice models and a Social Force model, thereby preserving both macro-level routing and fine-grained local crowd interactions.
\item We provide empirical validation using Wakayama Castle Park data collected through a field experiment covering both the park and adjacent public roads: the MLP model reproduces per-area temporal population trajectories with cosine similarity values exceeding 0.7 and reproduces mobility-induced distribution changes with a cosine similarity of 0.664.
\end{itemize}

%% file: 03_proposed_method_rewrite.tex
\section{Proposed Method}
\label{sec:proposed_method}

\subsection{Overview of the Proposed Method}
\color{black}

\begin{figure*}[t]
    \centering
    \includegraphics[width=0.85\textwidth]{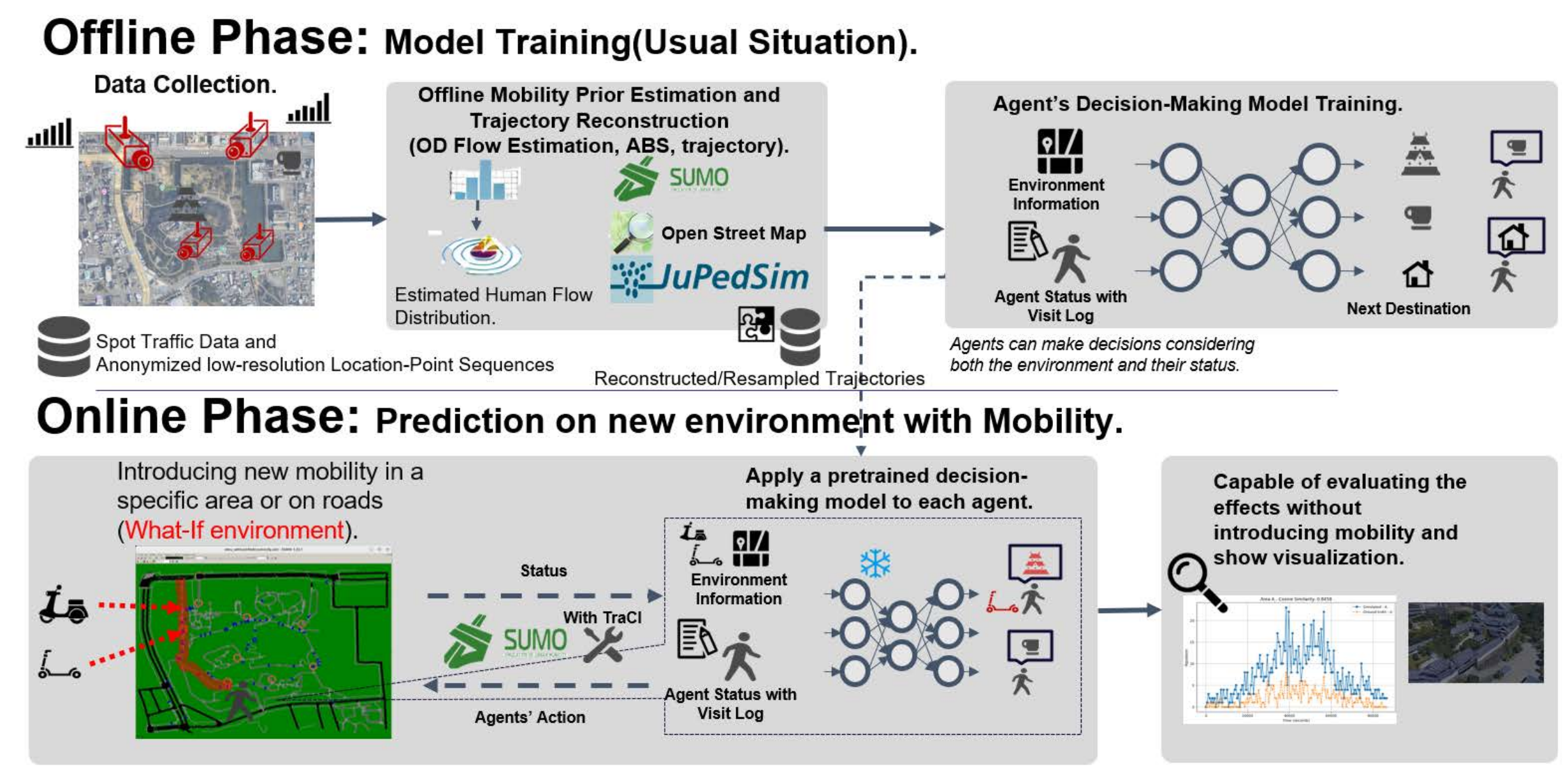}
    \caption{Overview of the Proposed Platform.}
    \label{fig:overview}
\end{figure*}

We propose a platform that predicts and visualizes how visitor behavior and aggregate flows change after the mobility introduction. 
Figure~\ref{fig:overview} provides an overview of the pipeline. 
In this study, we focus on low-speed, single-occupant personal mobility devices (e.g., electric kick scooters and small electric carts) intended for intra-site circulation. 
These devices are modeled as operating on a set of predefined mobility links that connect fixed segment endpoints; riders may board or alight only at these designated endpoints. 
The platform takes as input (i) normal-state anonymized/pseudonymized wide-area location-point sequences with tracking IDs and timestamps, (ii) time-stamped pedestrian traffic volumes measured at selected counting spots (counters), and (iii) environmental information such as the walkable network, Points of Interest (PoIs), and indicators of mobility introduction.
The output is a simulation scenario and a synthesized set of agents’ trajectories whose aggregate characteristics are consistent with the measured spot volumes. 
The resulting trajectories are mapped onto a 3D map using Unity and Cesium to support geographic inspection of simulation outcomes. 
Agents are instantiated to match the reconstructed departure events. 
At each decision step, an agent’s next-destination is updated by a destination-choice model, and the agent is then advanced in the simulator accordingly, yielding an agents’ trajectory. 
Each decision model maps an agent’s internal state (e.g., current location, time of day, and accumulated travel distance) and environmental inputs (e.g., pairwise PoI distances, PoI attractiveness, and movement costs) to a probability distribution over candidate next destinations. 
By switching the environmental inputs to represent mobility introduction (e.g., reduced generalized travel cost or increased attractiveness at mobility-enabled locations), the platform reproduces how agents’ choices adapt and how these adaptations aggregate into population-level flow changes.

\color{black}

\color{black}
\subsection{\textcolor{black}{Offline Phase: Data Preparation from Wide-Area Location Data and Spot Count}}
\label{sec:trajectory_reconstruction}

\subsubsection{Observation Model for Spot Counts and Wide-Area Location Sequences}
\label{sec:counts_collection}

\textcolor{black}{Following the pedestrian flow generation method of Uegaki et al.~\cite{ueg2023sim}, we introduce a structured movement prior estimated from wide-area location point sequences.}
\textcolor{black}{Wide-area location data refers to privacy-preserving location-point sequences collected from mobile devices or location-based services across a broad spatial region.
Such data are typically anonymized, coarsened in space or time, and sometimes perturbed or filtered to protect user privacy.
Consequently, they mainly describe aggregate inter-area movement tendencies rather than exact agents’ trajectories.}

The proposed pipeline combines normal-state wide-area location-point sequences with pedestrian traffic volumes measured at a subset of areas within a region of interest.
Let $\mathcal{A}$ denote the set of all areas covering the region of interest, and let $\mathcal{A}^{\mathrm{obs}}\subseteq\mathcal{A}$ denote the subset of areas where spot counts are available.
The location-point sequences are map-matched to the area network, yielding origin--destination observations over all ordered area pairs under normal conditions, whereas the spot counts provide accurate site-specific constraints on local flow intensity at $\mathcal{A}^{\mathrm{obs}}$.

Measurements are aggregated into discrete time slots; let $\mathcal{T}=\{1,2,\ldots,T\}$ be the set of slots, let $\Delta$ be the slot duration (e.g., $\Delta=10$ minutes), and denote the interval of slot $t\in\mathcal{T}$ by $I_t=[(t-1)\Delta,\; t\Delta)$.

For each observed area $a\in\mathcal{A}^{\mathrm{obs}}$ and slot $t\in\mathcal{T}$, we record the number of pedestrians passing the counting spot located in area $a$ during $I_t$, denoted by $L_{a,t}$.
Formally, letting $e_{i,a}(\tau)$ indicate that pedestrian $i$ passes the counting spot in area $a$ at time $\tau$, we define
\begin{equation}
L_{a,t}
=
\sum_{i} \mathbf{1}\!\left[\exists\,\tau \in I_t \ \text{s.t.}\ e_{i,a}(\tau)=1\right],
\label{eq:counts_def}
\end{equation}
where $\mathbf{1}[\cdot]$ is the indicator function.
Eq.~\ref{eq:counts_def} is a conceptual definition; in practice, we observe only the aggregated counts $L_{a,t}$ without identity-level linkage.
This yields a time-stamped dataset $\{L_{a,t}\}_{a\in\mathcal{A}^{\mathrm{obs}},\,t\in\mathcal{T}}$, which is used in the subsequent flow reconstruction module.

In addition, the wide-area location-point sequences are map-matched to obtain OD observations at the area level.
Let $(o_i,d_i)$ denote the origin and destination areas of movement sample $i$ after map matching, and let $v_i\in V$ denote the speed category assigned to that trajectory.
We define the observed OD counts as
\begin{equation}
X_{m,n,t,v}
=
\sum_i
\mathbf{1}\!\left[
o_i=m,\,
d_i=n,\,
\tau_i\in I_t,\,
v_i=v
\right],
\label{eq:od_obs}
\end{equation}
where $\tau_i$ is the departure time of movement sample $i$.
\color{black}
This yields a time-stamped OD dataset
$\{X_{m,n,t,v}\}_{m,n\in\mathcal{A},\,t\in\mathcal{T},\,v\in V}$.
The same map matched trajectory samples are also used before aggregation to estimate the continuous departure time and travel duration distributions described in Sec.~\ref{subsubsec:reconstruction}.
In the reconstruction stage, the aggregated counts provide the empirical relative OD composition.
Specifically, for each ordered area pair $(m,n)$, we define
\begin{equation}
c_{m,n}
=
\sum_{t\in\mathcal{T}}
\sum_{v\in V}
X_{m,n,t,v},
\label{eq:cmn_from_x}
\end{equation}
which represents the empirical number of observed transitions from area $m$ to area $n$ in the auxiliary location dataset.
\color{black}

\subsubsection{Flow Reconstruction and Calibration}
\label{subsubsec:reconstruction}
Given the spot-volume observations $\{L_{a,t}\}$ defined in Sec.~\ref{sec:counts_collection}, together with auxiliary normal-state mobility observations used to estimate inter-area priors, we reconstruct a statistical departure process and synthesize trajectories consistent with these aggregate measurements.
\textcolor{black}{We use a prior-demand generation and scaling procedure.}
Following prior count-based pedestrian-flow reconstruction methods~\cite{ueg2023sim}, we estimate plausible pedestrian departures while acknowledging the non-uniqueness of the reconstruction.


\color{black}
Because the spot counts provide only partial aggregate observations of the full pedestrian flow, they are insufficient to determine all area-to-area departures by themselves.
We therefore first construct a structured prior demand scenario from the wide-area location data and then calibrate its absolute scale using the observed spot counts.
\color{black}


\color{black}
For each ordered area pair $(m,n)\in \mathcal{A}\times\mathcal{A}$, we model the joint distribution of departure time of day and travel duration using a 2D Gaussian mixture model (GMM):
\begin{align}
p(\mathbf{z} \mid m,n)
&= \sum_{k=1}^{K}\pi_k^{(m,n)}
   \mathcal{N}\!\bigl(
     \mathbf{z} \mid \boldsymbol{\mu}_k^{(m,n)},
     \boldsymbol{\Sigma}_k^{(m,n)}
   \bigr)
\nonumber \\
&:= G_{m,n}(\mathbf{z}),
\qquad
\mathbf{z} = [t^{\mathrm{dep}}\;\; d^{\mathrm{trav}}]^\top,
\label{eq:prior_gmm}
\end{align}
where $t^{\mathrm{dep}}$ denotes the departure time of day and $d^{\mathrm{trav}}$ denotes the travel duration in minutes.
Here, $K$ is the number of Gaussian components, and $\pi_k^{(m,n)}$ is the mixture weight of component $k$ for movements from area $m$ to area $n$, satisfying $\pi_k^{(m,n)}\geq 0$ and $\sum_{k=1}^{K}\pi_k^{(m,n)}=1$.
The term $\mathcal{N}(\mathbf{z}\mid \boldsymbol{\mu}_k^{(m,n)},\boldsymbol{\Sigma}_k^{(m,n)})$ denotes a bivariate Gaussian density over the two-dimensional vector $\mathbf{z}$.
The mean vector $\boldsymbol{\mu}_k^{(m,n)}\in\mathbb{R}^2$ represents the typical departure time and travel duration of component $k$, while the covariance matrix $\boldsymbol{\Sigma}_k^{(m,n)}\in\mathbb{R}^{2\times 2}$ captures their variances and correlation.
Thus, each component can represent a characteristic movement pattern, such as short-duration trips during one period of the day or longer-duration trips during another period.

In practice, the set of priors $\{G_{m,n}\}$ is estimated offline from the unaggregated map matched trajectory samples used to construct Eq.~\eqref{eq:od_obs}.
The aggregated OD counts $\{X_{m,n,t,v}\}$ provide the empirical transition counts $\{c_{m,n}\}$ in Eq.~\eqref{eq:cmn_from_x}, which preserve the relative OD composition during sampling.
The estimated distributions are then treated as fixed priors during calibration.

\color{black}

For each ordered area pair $(m,n)$, we draw $c_{m,n}$ independent samples $(t^{\mathrm{dep}}_i,d^{\mathrm{trav}}_i)\sim G_{m,n}$ to form candidate trips from $m$ to $n$.
Let $dist(m,n)$ be the network path length between $m$ and $n$.
For each sample, we compute the implied walking speed
\begin{equation}
v_i=\frac{dist(m,n)}{d^{\mathrm{trav}}_i},
\label{eq:speed}
\end{equation}
discretize $v_i$ into a speed category $v\in V$, and discretize $t^{\mathrm{dep}}_i$ into a time slot $t\in\mathcal{T}$ with slot length $\Delta$.
Aggregating sampled trips yields an initial demand scenario
\begin{equation}
S=\{D_{m,n,t,v}\mid m,n\in \mathcal{A},\; t\in\mathcal{T},\; v\in V\},
\label{eq:scenario}
\end{equation}
where $D_{m,n,t,v}$ denotes the number of departures from $m$ to $n$ in time slot $t$ with speed category $v$.

We then calibrate only the slot-wise absolute scale using counter observations:
\begin{align}
r_t
&=
\frac{
  \sum_{a\in \mathcal{A}^{\mathrm{obs}}} L_{a,t}
}{
  \sum_{a\in \mathcal{A}^{\mathrm{obs}}}
  \sum_{m\in \mathcal{A}_a}
  \sum_{n\in \mathcal{A}}
  \sum_{v\in V}
  D_{m,n,t,v}
},
\nonumber \\
D'_{m,n,t,v}
&=
\left\lfloor
  r_t\,D_{m,n,t,v}
\right\rfloor,
\label{eq:scaling}
\end{align}
where $\mathcal{A}_a\subseteq \mathcal{A}$ denotes areas whose induced flows can contribute to the count at area $a$ under the adopted routing/assignment rule.
This rescaling compensates for the mismatch between the absolute scale implied by $\{c_{m,n}\}$ and the observed counter volumes while preserving the relative OD/time pattern encoded by the prior.
It matches the aggregate volume level per time slot; finer location-wise discrepancies are handled later by simulation and routing constraints.

\color{blue}
\color{black}

From the refined scenario $S'=\{D'_{m,n,t,v}\}$, we construct a simulation-ready set of agent departure events.
For each tuple $(m,n,t,v)$, we instantiate a total of $D'_{m,n,t,v}$ agents and assign each agent an origin--destination pair, 
a within-slot departure offset, and a walking-speed category consistent with $v$.
This yields a network-feasible continuous-time initialization of the calibrated macroscopic scenario, which is then passed to the microscopic simulation layer described next.

\subsubsection{Trajectory Refinement via Multi-Agent Simulation}
\label{subsubsec:traj_refine_multiagent}

The calibrated macroscopic scenario obtained in Sec.~\ref{subsubsec:reconstruction} provides a simulation-ready specification of area-to-area departures, but it is still coarse because it does not represent continuous pedestrian motion or local route choices within each area.
We therefore refine the reconstructed area-level flows with a microscopic multi-agent simulation layer.
The refined trajectories are then used to extract PoI-level decision events for training the destination-choice model.

To avoid confusion with the discrete time-slot index $t\in\mathcal{T}$, we use $s$ to denote continuous simulation time.
Let agent $i$ have state $(\mathbf{x}_i(s),\mathbf{u}_i(s),g_i(s))$, where $\mathbf{x}_i(s)\in\mathbb{R}^2$ is position, $\mathbf{u}_i(s)\in\mathbb{R}^2$ is velocity, and $g_i(s)$ is the current navigation target.
The kinematics are
\begin{equation}
\dot{\mathbf{x}}_i(s)=\mathbf{u}_i(s).
\end{equation}
We adopt the Social Force model to incorporate goal-directed motion and interactions:
\begin{equation}
m_i\,\dot{\mathbf{u}}_i(s)
=
m_i\frac{u^{\mathrm{max}}\mathbf{e}_i(s)-\mathbf{u}_i(s)}{\tau_i}
+
\sum_{j\neq i}\mathbf{f}_{ij}(s)
+
\sum_{o\in\mathcal{O}}\mathbf{f}_{io}(s),
\label{eq:social_force}
\end{equation}
where $m_i$ is an inertial scaling, $\tau_i$ is a relaxation time, $\mathbf{e}_i(s)$ is the unit vector pointing toward $g_i(s)$, $\mathbf{f}_{ij}(s)$ is the interaction force from another agent $j$, and $\mathbf{f}_{io}(s)$ is the interaction force from obstacle $o\in\mathcal{O}$.
\color{black}
Here, $u^{\mathrm{max}}$ denotes the simulator-assigned upper bound on the agent's admissible speed.
This speed cap is determined by the current simulation context and may vary over time according to the agent's mobility mode.
\color{black}
The dynamics are simulated with a fixed integration time step supported by the underlying simulator.

We implement this refinement by coupling SUMO~\cite{Dan2002sumo} and JuPedSim~\cite{arm2015juped}.
SUMO enforces network feasibility (walkable topology, crossings, and signal control) and executes macroscopic routing, while JuPedSim provides microscopic pedestrian dynamics based on the Social Force model~\cite{arm2015juped,helbing1991mathematical}.
This coupling yields continuous trajectories that remain consistent with the reconstructed area-level scenario while becoming locally plausible under congestion through interaction-aware motion.

To construct PoI-level decision data, we detect when a refined trajectory enters the vicinity of a PoI.
Let $\mathcal{P}$ denote the set of PoIs used as candidate destinations and mobility endpoints.
Each PoI $p\in\mathcal{P}$ has a spatial coordinate $\mathbf{r}_p\in\mathbb{R}^2$ and we regard agent $i$ as visiting PoI $p$ when
\begin{equation}
\|\mathbf{x}_i(s)-\mathbf{r}_p\|\le \rho_p,
\end{equation}
where $\rho_p$ is the PoI-specific vicinity radius.
At each such event, we record the current agent state, the local environment, and the next destination PoI.

\subsection{Destination-Choice Model Design}

The objective of the destination-choice model is to reproduce reconstructed movement trajectories. Because observational data are typically available for only one of the two conditions (with or without mobility introduction) under non-identical settings, it is difficult to directly measure how overall flows would change if a mobility service were introduced. 
Instead, we learn how visitors respond to environmental factors from observed data, and then treat mobility introduction as an environmental modification applied to the learned decision model.

Concretely, whenever an agent appears or arrives at a destination the decision model observes the current environment and the agent’s internal state, then selects the next destination. 
The process repeats until either the decision model chooses to exit or the agent's stamina—which is decremented after every movement—reaches zero.
The environmental inputs to the decision model include current time, PoI attractiveness values, and pairwise distances between PoIs. 
The agent-state inputs include the current location, previously visited PoIs, and cumulative walking distance

To implement this framework, we parametrize the decision model with neural networks and evaluate a set of candidate architectures. 
We consider two deterministic models: a multi-layer perceptron (MLP), which treats the input features as a fixed-length vector, and a graph neural network (GNN), which propagates information over the PoI graph through message passing. 
These variants are used to examine whether explicitly modeling PoI nodes and pairwise relations improves prediction accuracy. 
In contrast, we also consider a probabilistic variant, MLP + Mixture-of-Softmax (MoS), which learns a mixture distribution over candidate destinations and captures multi-modal choice behavior. 
During simulation, the next destination is sampled from this output distribution so that agents' trajectories are stochastic and reflect decision uncertainty.

\subsection{Online Phase: Mobility-Aware Simulation}
\label{sec:online_phase}

In the online phase, agent behavior is generated by applying the trained destination-choice model within the simulator under a modified environment that reflects mobility introduction.
Let $f_{\Theta}$ denote the trained decision model, and let $s_i(t)$ denote the current state of agent $i$ at decision epoch $t$, and $\phi(t)$ denote the current environmental features.
For deterministic models, \(q_i^{next}=\arg\max_q f_\Theta(q\mid s_i(t),\phi(t))\); for probabilistic models, \(q_i^{next}\sim f_\Theta(q\mid s_i(t),\phi(t))\).

To represent mobility introduction, we do not retrain the decision model on intervention-specific data.
Instead, we modify the environmental features supplied to the trained model.
Specifically, for mobility-enabled PoI pairs and affected PoI locations, we replace the effective travel speed with the mobility cruising speed and adjust the attractiveness of associated locations.
\color{black}
The simulator-assigned upper bound on the agent's admissible speed on link $(p,q)$ is defined as
\begin{equation}
\tilde{u}_{p,q}^{max} =
\begin{cases}
u^{\mathrm{max_{mob}}}, & \text{mobility is available between $p$ and $q$},\\
u^{\mathrm{max_{walk}}}, & \text{otherwise},
\end{cases}
\end{equation}
and the corresponding pairwise distances between PoIs $\tilde{d}_{p,q}$ as
\begin{equation}
\tilde{d}_{p,q} = \frac{dist(p,q)}{\tilde{u}_{p,q}^{max}}.
\end{equation}
Similarly, for locations affected by the introduction of mobility, attractiveness is updated. 
\color{black}

These modified environmental variables are then fed into the trained decision model in place of the baseline features.
As a result, the model reproduces how agents adapt their destination choices under mobility introduction without requiring a new intervention-specific training dataset.
Operationally, agents are spawned according to the reconstructed departures, propagated on the network by SUMO/JuPedSim, and their targets or route segments are updated via TraCI whenever a new destination decision is triggered.
This online procedure yields counterfactual agent trajectories under mobility introduction while preserving the macroscopic departure schedule estimated in the offline phase.

\color{black}

\subsection{Visualization of Simulation Results}

\begin{figure}[t]
    \centering
    \includegraphics[width=\linewidth]{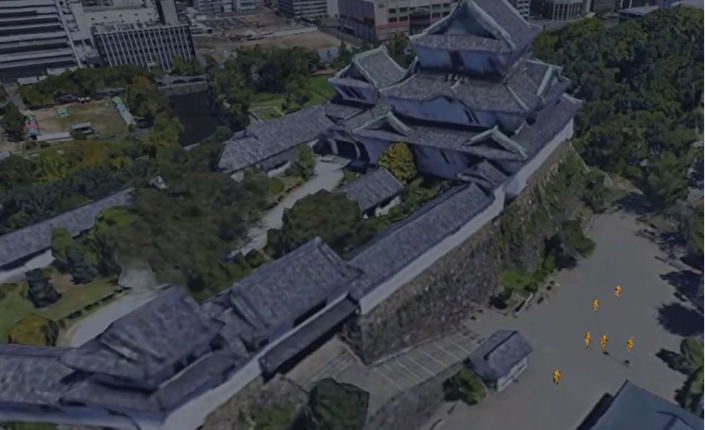}
    \caption{Example of Visualized Result.}
    \label{fig:3d-2}
\end{figure}

To facilitate intuitive understanding of mobility effects, we visualize simulated agent positions in three-dimensional space. 
During a SUMO run, TraCI enables sequential retrieval of each agent’s ID and location; by accumulating this time-series data, we can trace agents’ trajectories. 
We render these trajectories in Unity on top of Cesium’s 3D map tiles, placing agents according to the time-stamped positions obtained from TraCI. 
An example of the visualization output is shown in Fig.~\ref{fig:3d-2}; the rendered scene depicts Wakayama Castle Park, with agents rendered in yellow traversing the lower-right portion of the view. This visualization aids the interpretation of spatio-temporal changes in pedestrian flow by making local increases and decreases in movement visually apparent.

%% file: 04_experiment_alter.tex
\section{Experimental Evaluation}
\label{sec:experiment_and_result}

\subsection{Objectives and Evaluation Metrics}

\begin{figure*}[t]
  \centering

  \begin{subfigure}[t]{0.23\textwidth}
    \centering
    \includegraphics[height=2.5cm,keepaspectratio]{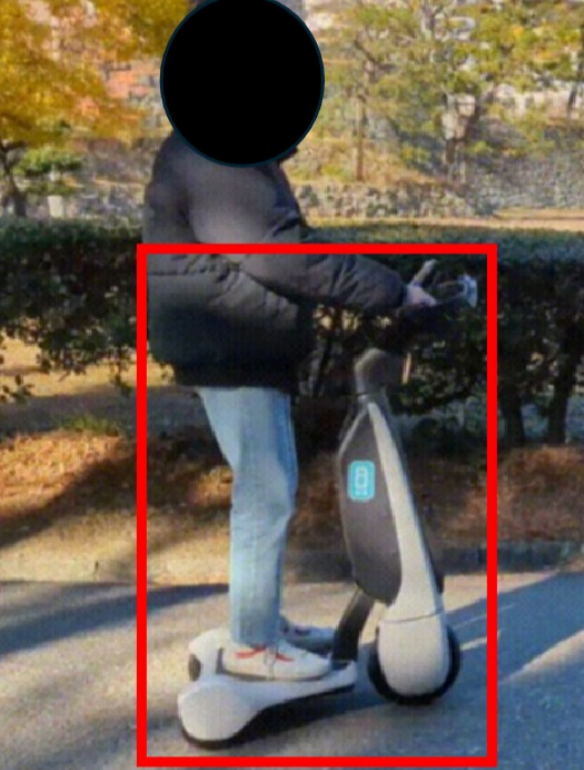}
    \caption[C+walk T]{C\textsuperscript{+}walk$_{\mathrm{T}}$\\
    {\scriptsize three-wheeled standing-type battery electric vehicle (BEV).}\protect\footnotemark}
    \label{subfig:mobility1}
  \end{subfigure}
  \hfill
  \begin{subfigure}[t]{0.23\textwidth}
    \centering
    \includegraphics[height=2.5cm,keepaspectratio]{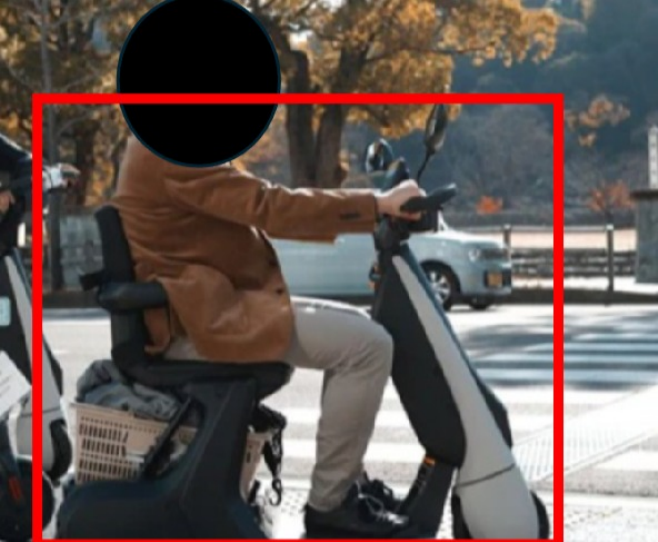}
    \caption[C+walk S]{C\textsuperscript{+}walk$_{\mathrm{S}}$\\
    {\scriptsize pedestrian-mobility-assistance model.}\protect\footnotemark}
    \label{subfig:mobility2}
  \end{subfigure}
  \hfill
  \begin{subfigure}[t]{0.23\textwidth}
    \centering
    \includegraphics[height=2.5cm,keepaspectratio]{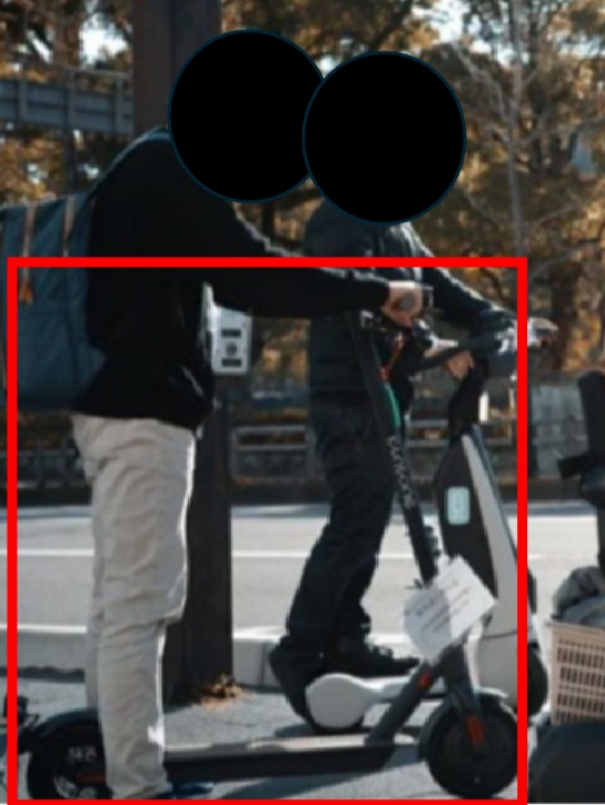}
    \caption[Kintone Model One S]{Kintone Model One~S\\
    {\scriptsize electric kick scooter.}\protect\footnotemark}
    \label{subfig:mobility3}
  \end{subfigure}
  \hfill
  \begin{subfigure}[t]{0.23\textwidth}
    \centering
    \includegraphics[height=2.5cm,keepaspectratio]{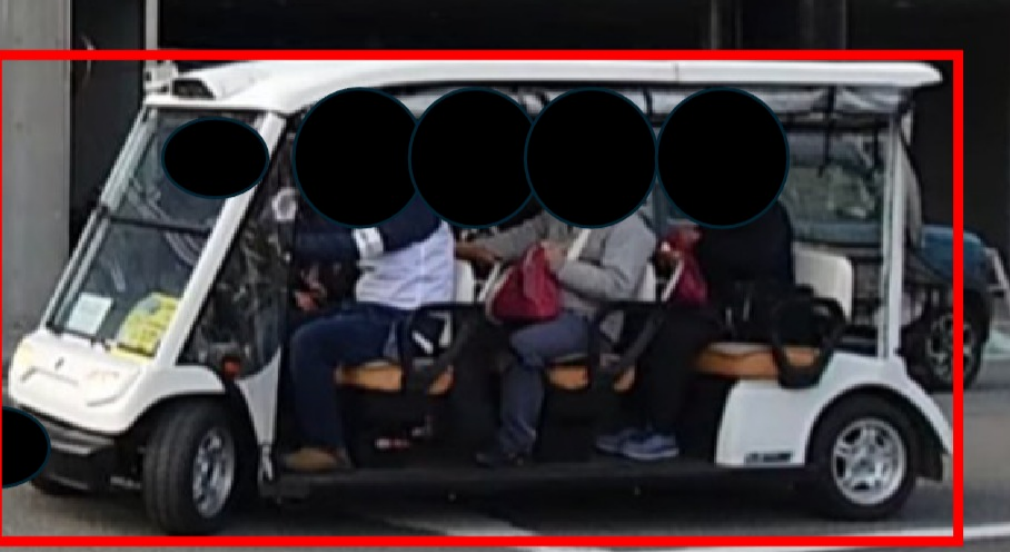}
    \caption[Yamaha AR-07 Green Slow Mobility]{Yamaha AR-07 Green Slow Mobility\\
    {\scriptsize low-speed electric vehicle.}\protect\footnotemark}
    \label{subfig:mobility4}
  \end{subfigure}

  \caption{\textcolor{black}{Mobilities we installed.}}
  \label{fig:mobilities}
\end{figure*}





We evaluate the proposed simulation platform by quantifying its ability to reproduce observed spatial and temporal patterns of pedestrian presence in the study area, using population distributions derived from field observations around Wakayama Castle Park. 
The evaluation pursues two principal aims: first, to assess the overall fidelity of simulated population distributions and their temporal dynamics; and second, to determine how specific modelling choices affect that fidelity. 
As primary quantitative metrics we use mean absolute error (MAE) computed across partitioned areas (and its day-aggregated sum) to measure spatial-count accuracy, and cosine similarity to assess agreement of temporal trajectories. 

This study is organized around four research questions.
First, to what extent the simulator reproduces observed spatial and temporal population distributions under mobility.
Second, how the choice of destination decision model architecture, such as a multilayer perceptron or a graph neural network, influences reproduction quality and spatial error patterns, particularly in regions prone to spurious congestion.
Third, whether probabilistic decision making improves aggregate prediction accuracy or uncertainty characterization compared with deterministic argmax selection.
Fourth, which exit mechanism better reproduces empirical exit time distributions and their downstream effects on population dynamics.

\subsection{Dataset and Preprocessing}
\label{subsec:dataset}
\color{black}
Field observations were acquired within a 1 km$^2$ area centered on Wakayama Castle Park.
Two sensing campaigns were carried out.
The first campaign collected wide-area GPS location records continuously over one month (31 days) to capture typical, normal-state mobility patterns.
These records were used as the wide-area location dataset for the GMM-based simulation procedure described in Sections~\ref{sec:counts_collection} and \ref{subsubsec:reconstruction}, and to generate the baseline trajectory ensemble without mobility, which was used to construct the destination-choice training dataset.
The second campaign was a separate seven-day field experiment in which high-resolution pedestrian spot counts were recorded with LiDAR sensors at preselected counting locations.
These LiDAR measurements were used as spot-volume observations for calibrating the reconstructed flows, as described in Section~\ref{subsubsec:reconstruction}.
\color{black}
All spot measurements were aggregated into 10-minute intervals for downstream processing.
All synthetic trajectories and training targets used in this work were generated using the method described in Section \ref{sec:counts_collection}, \ref{subsubsec:reconstruction}, and \ref{subsubsec:traj_refine_multiagent}. 
From the synthesized trajectory ensemble, we produce three derived data products that serve distinct purposes in the experiments. Time- and location-specific arrival (spawn) rates and per-area, per-time-step population time series were generated for both the baseline (no mobility introduced) and the mobility-introduced conditions — the mobility-introduced versions were obtained by calibrating reconstructed flows to the seven days of LiDAR observations collected during the intervention week. The destination-choice training dataset (decision-epoch records paired with the selected next destination, including state and spatial features such as current PoI, time of day, cumulative travel distance, and distances to candidate PoIs) was created only from the baseline (without mobility) reconstructed trajectories and used to train the decision models under normal-state behavior assumptions.

\subsection{Simulation Settings}
We performed simulations covering a single day (86,400 seconds, 24 hours) for Wakayama Castle Park in Japan. 
The park geometry and road network were reconstructed from OpenStreetMap (OSM) data.

\subsubsection{Area Partitioning and Attraction Settings}
\label{subsec:areas_attraction}

\begin{figure}[t]
    \centering
    \includegraphics[width=\linewidth]{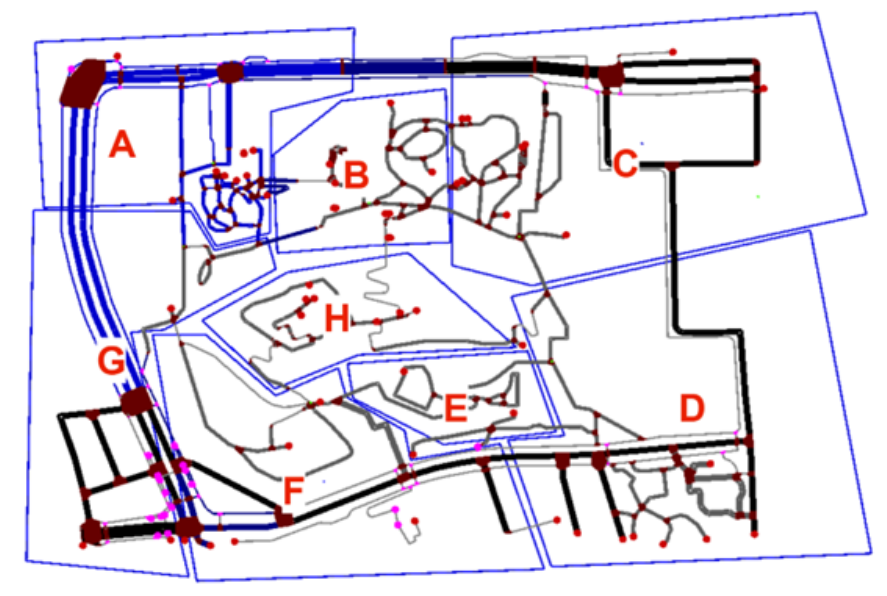}
    \caption{Area Partitioning for our Simulation.}
    \label{fig:taz-1}
\end{figure}

\begin{figure}[t]
    \centering
    \includegraphics[width=\linewidth]{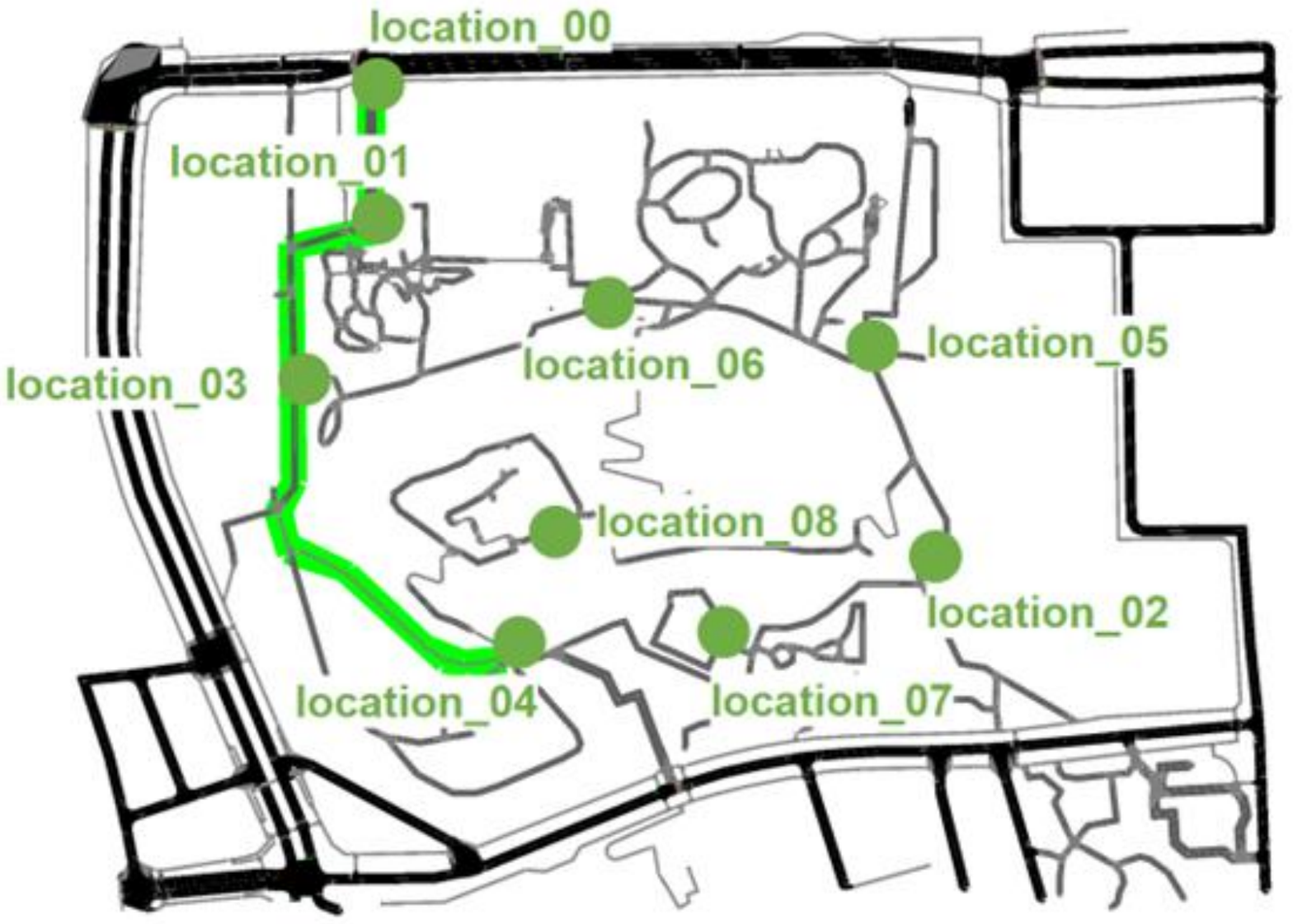}
    \caption{PoI Locations and Mobility Deployment.}
    \label{fig:loc}
\end{figure}

\begin{table}[t]
\centering
\caption{PoI Attraction Values Without and With Mobility Introduction.}
\begin{tabular}{|c|c|cc|}
\hline
\multirow{2}{*}{PoI ID} & \multirow{2}{*}{Area} & \multicolumn{2}{c|}{Attraction Value}             \\ \cline{3-4}
&                      & \multicolumn{1}{c|}{Without Mobility} & With Mobility \\ \hline
00                    & A                    & \multicolumn{1}{c|}{0.0680}    & 0.0545  \\ \hline
01                    & A                    & \multicolumn{1}{c|}{0.0680}    & 0.0545  \\ \hline
02                    & D                    & \multicolumn{1}{c|}{0.157}   & 0.151   \\ \hline
03                    & G                    & \multicolumn{1}{c|}{0.138}   & 0.153   \\ \hline
04                    & F                    & \multicolumn{1}{c|}{0.140}     & 0.152   \\ \hline
05                    & C                    & \multicolumn{1}{c|}{0.00950}   & 0.0990   \\ \hline
06                    & B                    & \multicolumn{1}{c|}{0.0962}   & 0.0978   \\ \hline
07                    & E                    & \multicolumn{1}{c|}{0.0679}   & 0.0628   \\ \hline
08                    & H                    & \multicolumn{1}{c|}{0.165}   & 0.176  \\ \hline
\end{tabular}
\label{tab:attraction_store}
\end{table}


The park area was partitioned into areas used to define each pedestrian’s origin and destination areas within the simulation; population distributions are computed with respect to these areas. 
We defined specific landmarks (e.g., the zoo, the castle keep, parking lots) as PoIs that are expected to attract frequent visits. 
Each area contains at least one PoI. 
Visualizations of the area partitioning and PoI locations are shown in Fig.~\ref{fig:taz-1} and Fig.~\ref{fig:loc}.
Each PoI is assigned an attraction score representing the propensity of visitors to want to visit that location. 
Attraction scores were derived from the per-area visit proportions in the dataset described in Section~\ref{subsec:dataset} and normalized across all PoIs. 
If multiple PoIs are contained within a single area, the area’s visit proportion is split equally among those PoIs. 
Because the attraction scores are normalized over all PoIs, the score of a given location may decrease in relative terms with mobility introduction as the scores of other locations increase.
The mapping from PoIs to zones and the attraction values without/with mobility introduction used in the experiments are given in Table~\ref{tab:attraction_store}.

\subsubsection{Mobility Introduction Settings}
In accordance with the field experiment, mobility was introduced on the routes indicated in green in Fig.~\ref{fig:loc}.
For the experiments, we set the mobility vehicle cruising speed to 20 km/h and the pedestrian walking speed to 5.0 km/h. 
In addition, attraction scores were updated at deployment locations such as location\_00, location\_01, location\_03, and location\_04.
The micromobilities we installed are shown in Fig~\ref{fig:mobilities}.

\addtocounter{footnote}{-3}
\footnotetext{{\scriptsize Toyota Motor Corporation, ``Toyota Launches the C+walk T in Japan, a New Form of Walking-Area Mobility,'' \url{https://global.toyota/en/newsroom/toyota/35872225.html}.}}

\stepcounter{footnote}
\footnotetext{{\scriptsize Toyota Motor Corporation, ``Toyota Launches the C+walk S in Japan, a New Form of Walking-Assistance Mobility,'' \url{https://global.toyota/en/newsroom/toyota/38947604.html}.}}

\stepcounter{footnote}
\footnotetext{{\scriptsize Kintone official product page, ``Kintone Model One S,'' 

\url{https://kintone.mobi/products/modelone-s?variant=42886564970678}.}}

\stepcounter{footnote}
\footnotetext{{\scriptsize Yamaha Motor Co., Ltd., ``Low-Speed Electric Mobility for First- and Last-Mile Transportation,'' \url{https://global.yamaha-motor.com/news/2023/0825/newsletter.html}.}}

\subsection{Reproducing Human Flows With Mobility Introduction}


\begin{table}[t]
\centering
\caption{Population Reproduction Accuracy With Mobility Introduction.}
\begin{tabular}{|c|c|c|}
\hline
Model &  MAE & Day-aggregated MAE\\ \hline
MLP & 2.34 & 334\\ \hline
MLP + MoS & 2.52 & 360 \\ \hline
GNN & 2.87 & 410 \\ \hline
MLP with Exit Class & 1.76 & 252 \\ \hline
MLP + MoS with Exit Class & 1.75 & 250 \\ \hline
\end{tabular}
\label{tab:mae-sim-mobi}
\end{table}

\begin{figure*}[ht]
  \centering

  \begin{subfigure}{0.32\textwidth}
    \centering
    \includegraphics[width=\linewidth]{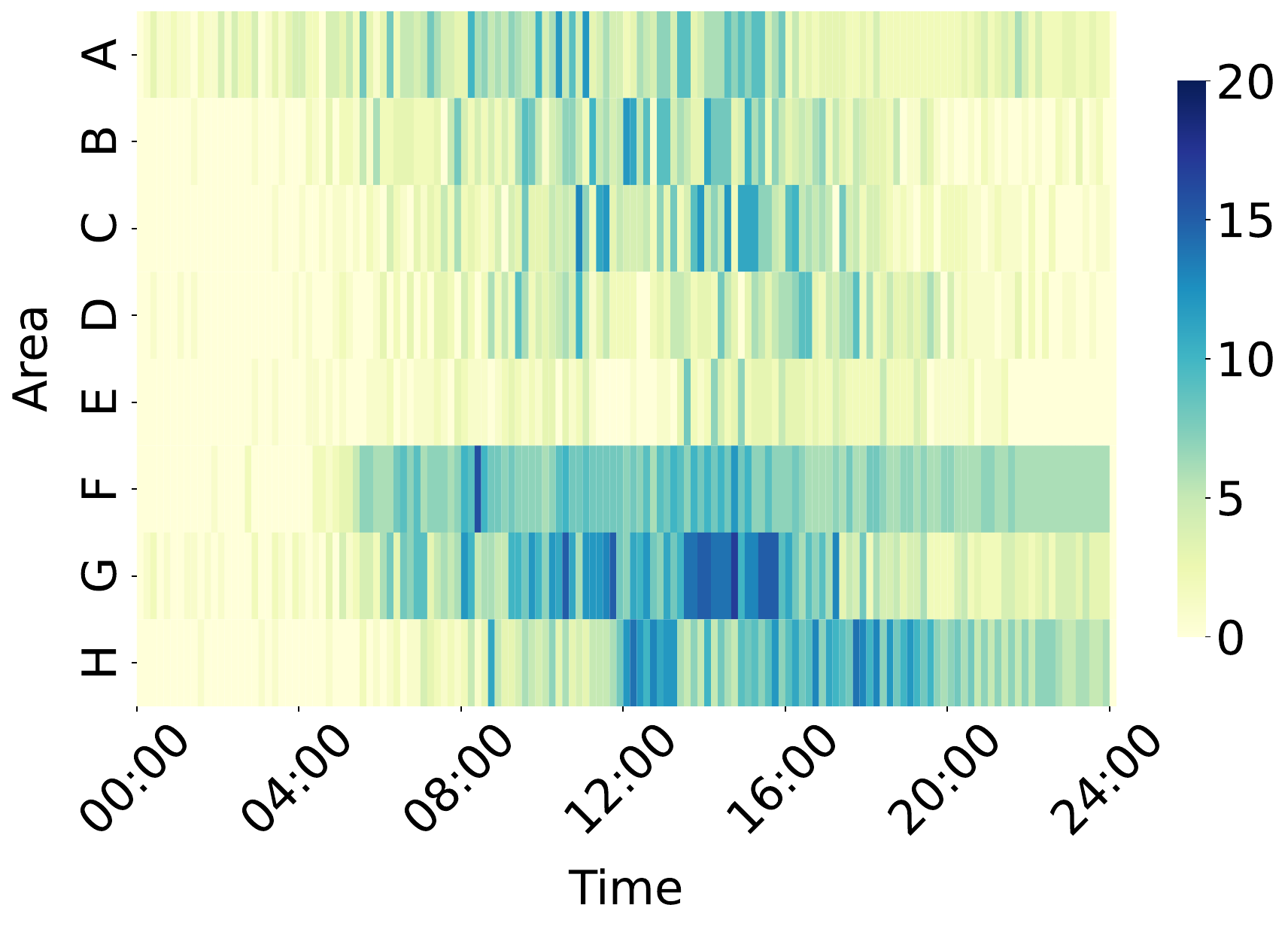}
    \caption{MLP Model}
    \label{subfig:mlp-dist}
  \end{subfigure}
  \hfill
  \begin{subfigure}{0.32\textwidth}
    \centering
    \includegraphics[width=\linewidth]{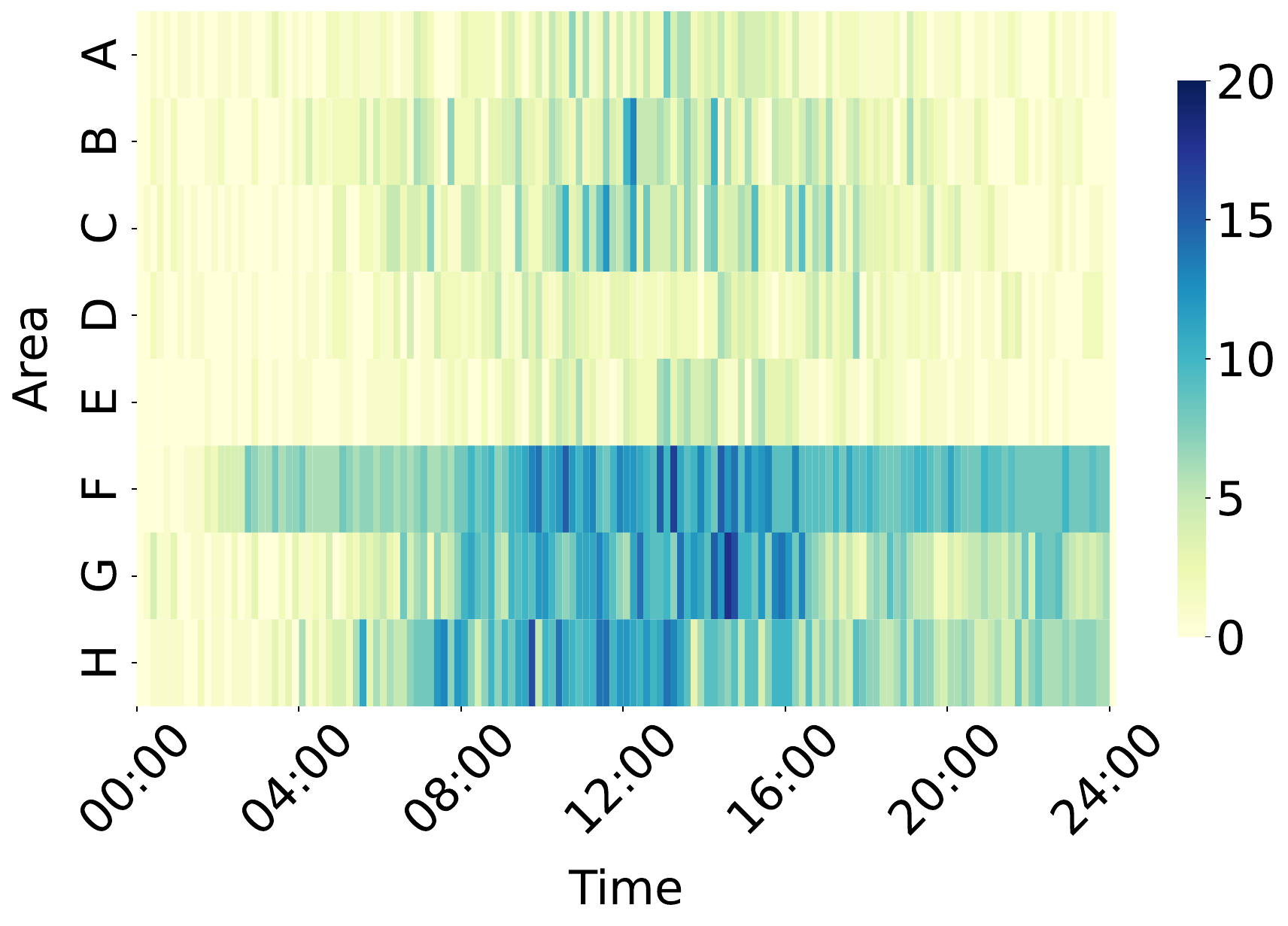}
    \caption{MLP+MoS Model}
    \label{subfig:mos-dist}
  \end{subfigure}
  \hfill
  \begin{subfigure}{0.32\textwidth}
    \centering
    \includegraphics[width=\linewidth]{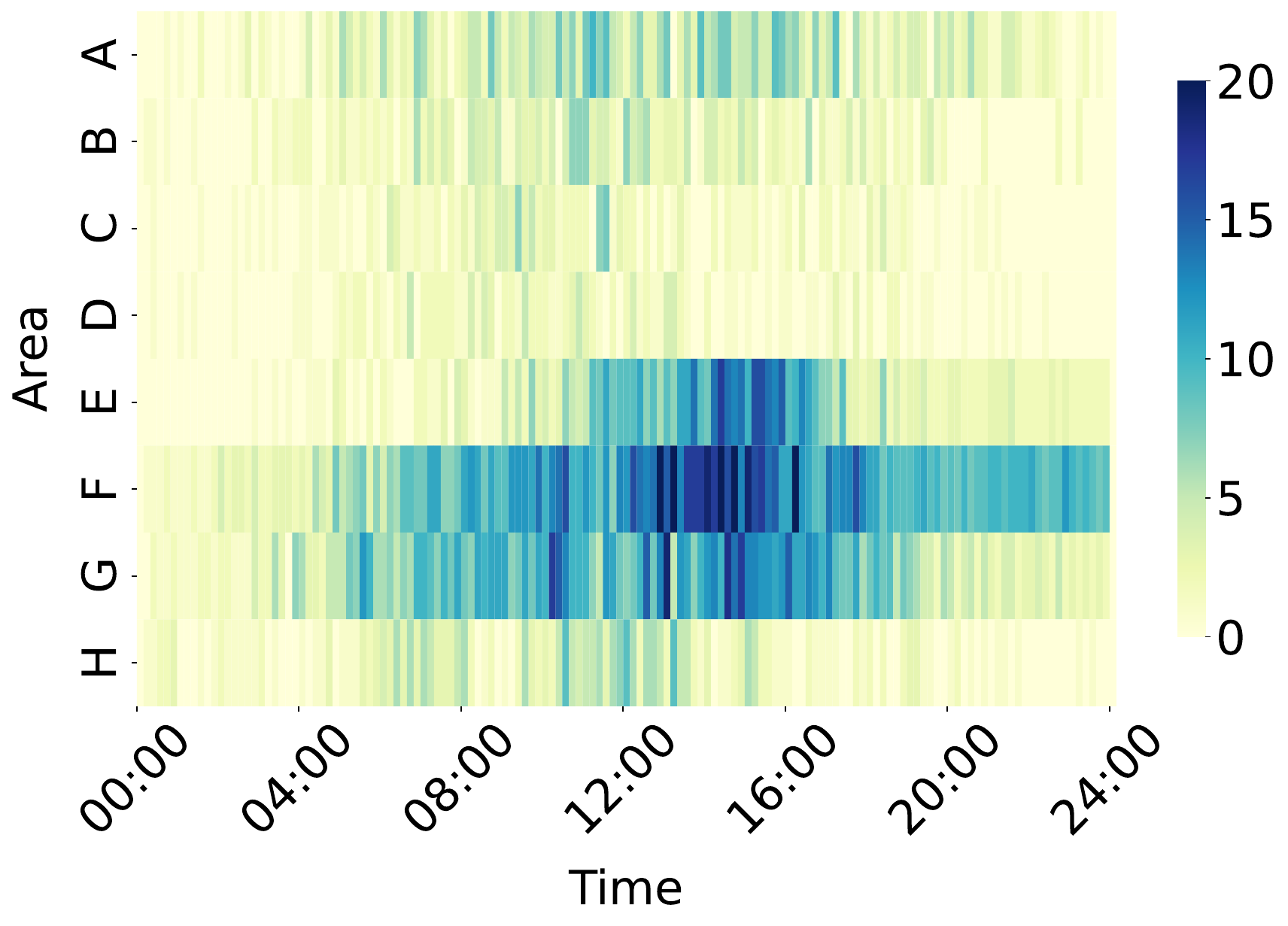}
    \caption{GNN Model}
    \label{subfig:gnn-dist}
  \end{subfigure}


  \begin{subfigure}{0.32\textwidth}
    \centering
    \includegraphics[width=\linewidth]{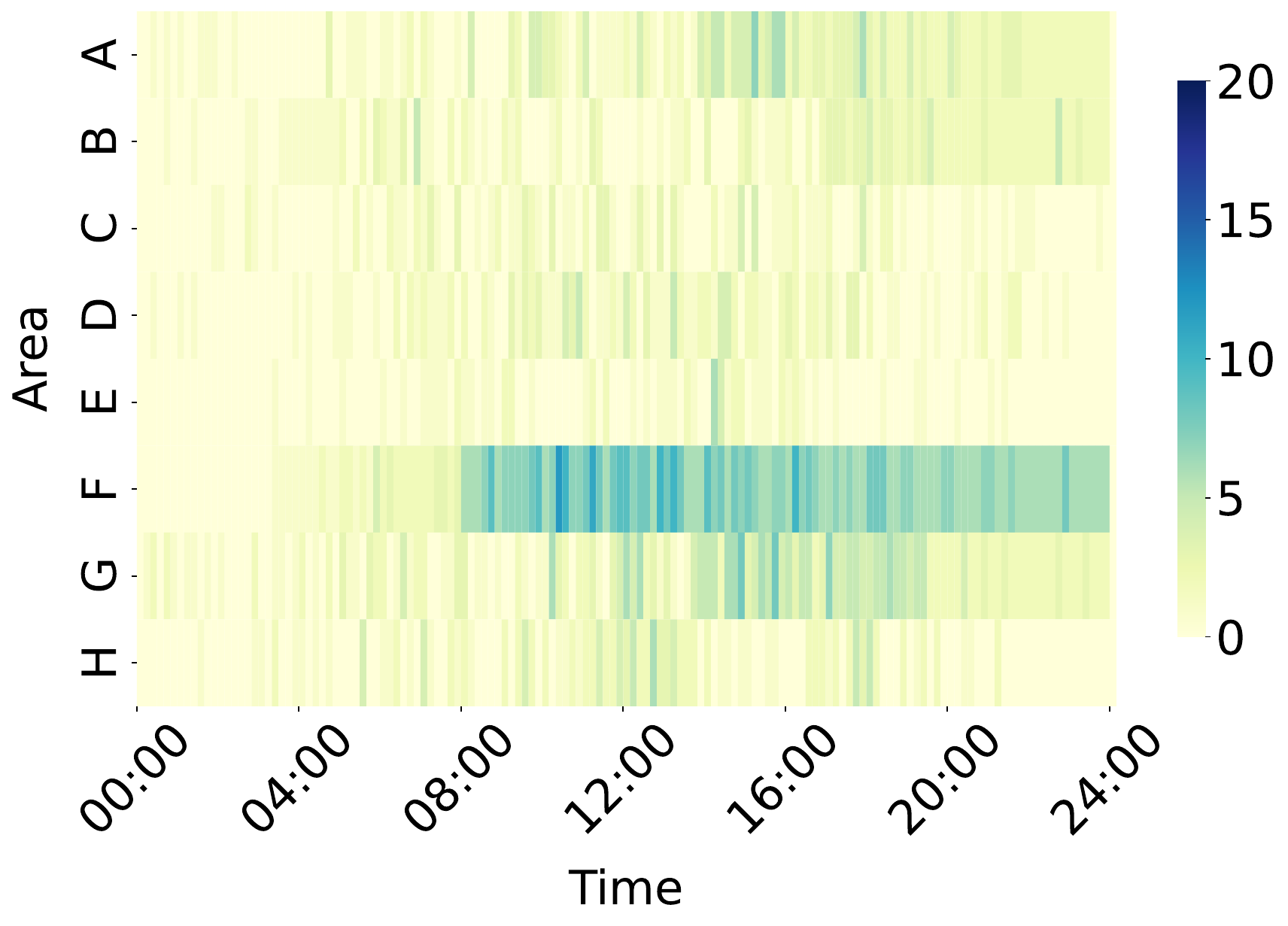}
    \caption{MLP Model with Exit Class}
    \label{subfig:model4}
  \end{subfigure}
  \hfill
  \begin{subfigure}{0.32\textwidth}
    \centering
    \includegraphics[width=\linewidth]{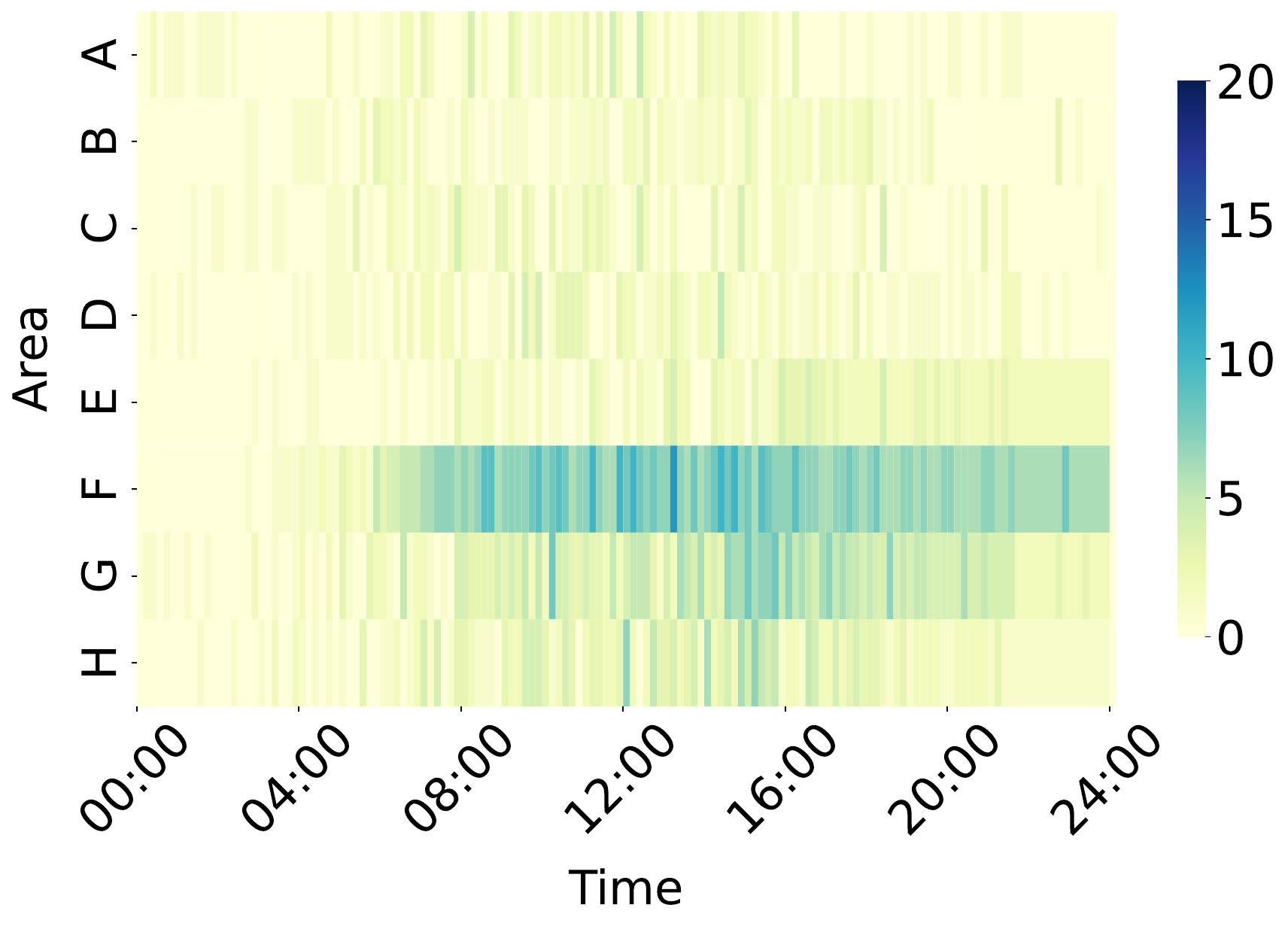}
    \caption{MLP+ MoS Model with Exit Class}
    \label{subfig:model5}
  \end{subfigure}
  \hfill
  \begin{subfigure}{0.32\textwidth}
    \centering
    \includegraphics[width=\linewidth]{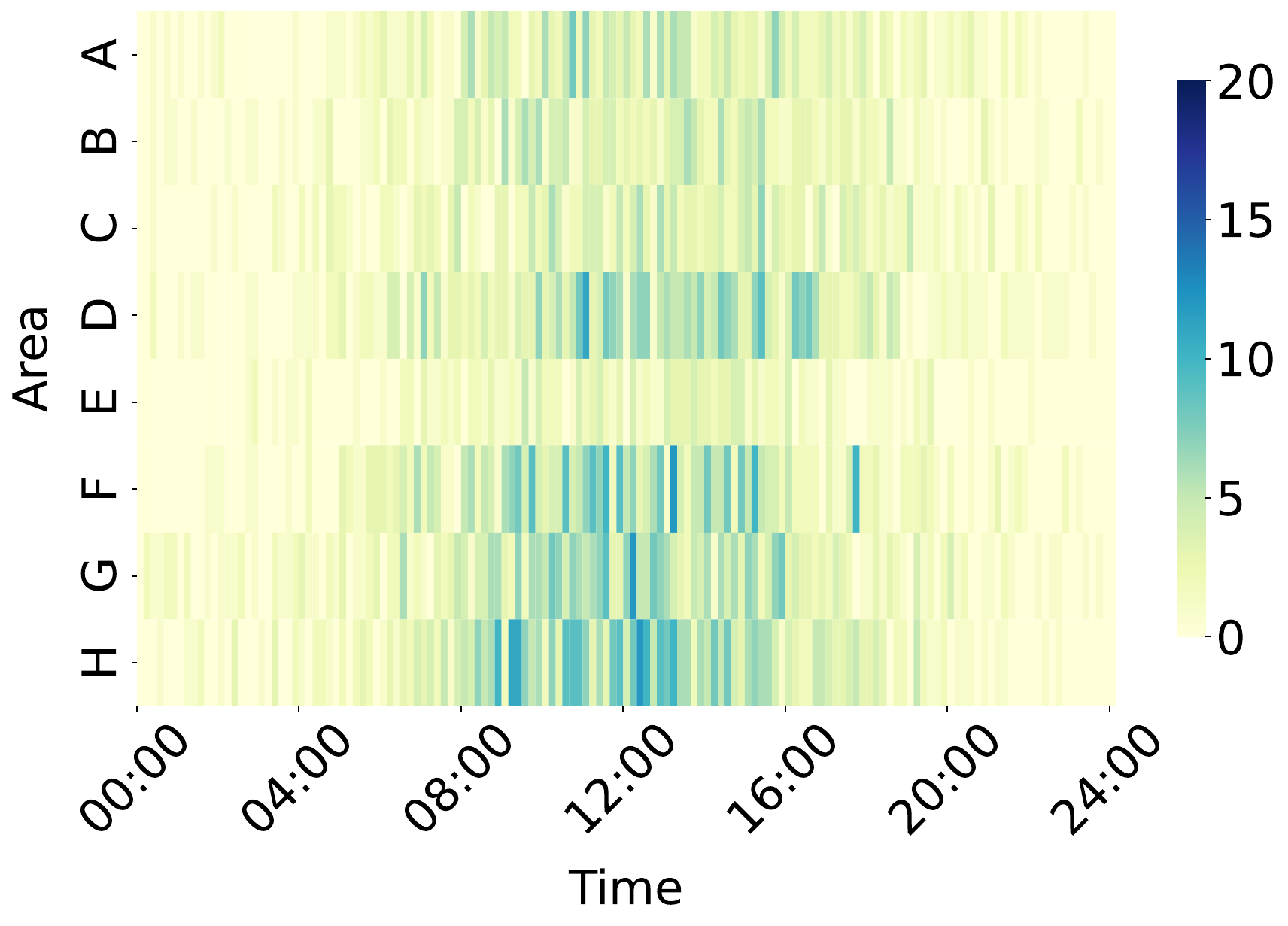}
    \caption{Ground Truth Distribution}
    \label{subfig:gt}
  \end{subfigure}

  \caption{Estimated and Ground Truth Population Distributions With Mobility Introduction.}
  \label{fig:distribution}
\end{figure*}

\begin{table}[t]
\centering
\caption{Cosine similarities for two evaluation criteria.}
\label{tab:cosine-combined}
\begin{tabular}{|l|c|c|}
\hline
Model & \shortstack{Population} & \shortstack{Change in Population} \\ \hline
MLP & 0.769 & 0.664 \\ \hline
MLP+MoS & 0.758 & 0.694 \\ \hline
GNN & 0.663 & 0.578 \\ \hline
MLP with Exit Class & 0.624 & 0.582 \\ \hline
MLP + MoS with Exit Class & 0.653 & 0.619 \\ \hline
\end{tabular}
\end{table}

\subsubsection{Overall Fidelity of the Proposed Platform}

We first evaluate the overall fidelity of the proposed simulation platform in reproducing observed pedestrian flows. 
Figure~\ref{fig:distribution} presents the simulated temporal population distributions per area together with the ground-truth distribution. 
Table~\ref{tab:mae-sim-mobi} summarizes the MAE and day-aggregated MAE for each model configuration, while Table~\ref{tab:cosine-combined} reports two cosine-similarity metrics: the similarity between simulated and observed population time-series under the mobility-enabled scenario, and the similarity of the mobility-induced change patterns computed as the difference between the mobility-enabled and baseline time-series.

Regarding spatial population reproduction, the day-aggregated MAE ranges from approximately 250 to 400 depending on the model configuration. 
Across all models, Regions F and G tend to be overestimated, suggesting that the simulator may slightly overpredict the increase in visitor concentration caused by the mobility introduction. 
Nevertheless, the cosine similarity results indicate that the simulator captures the overall temporal patterns of population dynamics reasonably well. 
For both the population trajectories and the mobility-induced changes, cosine similarities are ranging from 0.578 to 0.769, with the MLP reaching 0.769, indicating that the platform can reproduce the general temporal trends of pedestrian flows in the study area.

\subsubsection{Effect of Network Architecture}

Next, we investigate how the architecture of the destination-decision model affects the reproduction of human flows by comparing MLP and GNN models. 
As shown in Figure~\ref{fig:distribution}, only the GNN model produces a spurious congestion pattern in Area~E that does not appear in the ground-truth data. 
This behavior is also reflected in the larger MAE values reported in Table~\ref{tab:mae-sim-mobi}.

These results suggest that explicitly modeling spatial relationships between PoIs through a graph structure does not necessarily improve aggregate flow reproduction in this setting. 
Because the study area covers only approximately $1\,\mathrm{km}^2$, the spatial distances between areas are relatively small and walking remains a feasible option across most locations. 
Consequently, the benefits of modeling detailed spatial relationships through a GNN appear limited in this compact environment.

\subsubsection{Probabilistic vs.\ Deterministic Decision Modeling}

We further examine the effect of probabilistic decision modeling. 
Specifically, we compare a deterministic MLP model, which selects the next destination using an argmax rule, with an MLP augmented with a Mixture-of-Softmax (MoS) output layer that learns a distribution over candidate destinations and samples the next destination according to that distribution.

The results shown in Table~\ref{tab:mae-sim-mobi} and Table~\ref{tab:cosine-combined} indicate that there is no substantial difference in accuracy between the two approaches. 
Both the MAE of the reproduced population distribution and the cosine similarity of the temporal population trajectories are comparable between the deterministic and probabilistic models. 
This suggests that, under the present experimental conditions, modeling decision uncertainty through a probabilistic output distribution does not significantly improve the aggregate reproduction accuracy of pedestrian flows.

\subsubsection{Exit Mechanism and Its Impact}

Next, we compare two approaches for modeling visitors' exit behavior from the park. 
In the first approach, exiting the park is treated as an explicit class within the destination-choice model, allowing the exit timing to be learned directly as part of the classification task. 
In the second approach, exit behavior is modeled using a stamina variable that decreases as the agent moves; agents leave the simulation when the stamina value reaches zero, and the stamina parameters are tuned based on the training data.

In Figure~\ref{fig:distribution}, Table~\ref{tab:mae-sim-mobi}, and Table~\ref{tab:cosine-combined}, the models labeled with Exit Class correspond to the former approach, while the remaining models employ the stamina-based mechanism. 
Table~\ref{tab:mae-sim-mobi} shows that incorporating the exit event as an explicit class leads to lower MAE values, indicating improved accuracy in reproducing the spatial population distribution. 
In contrast, the cosine similarity results in Table~\ref{tab:cosine-combined} suggest that the stamina-based mechanism achieves slightly better agreement with the observed temporal population trajectories. 
These results indicate that each approach has distinct advantages, and the appropriate choice may depend on the evaluation objective of the simulation.

\subsubsection{Analysis of the Decision-making Model}
\color{black}
\begin{figure}
    \centering
    \includegraphics[width=\linewidth]{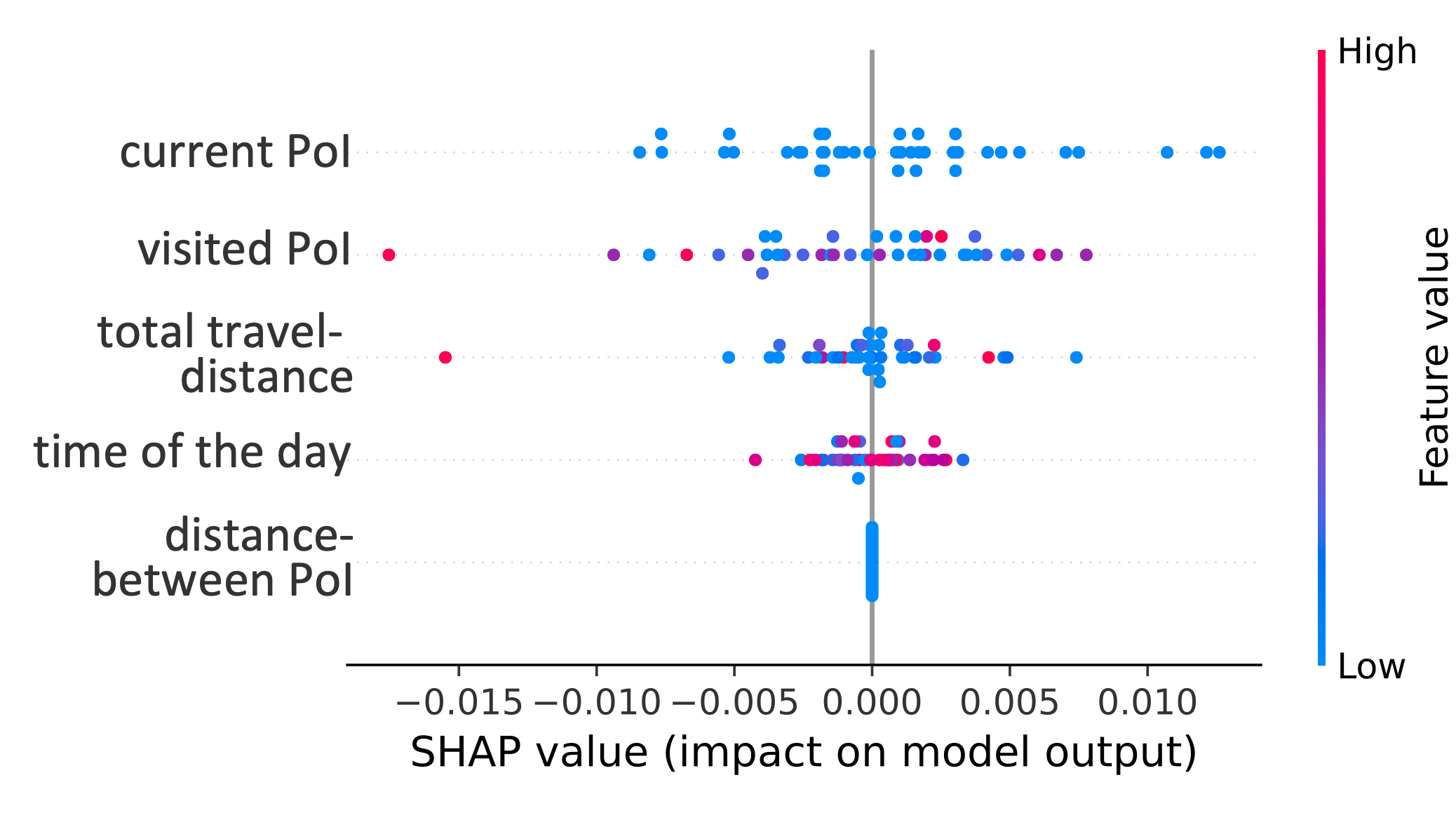}
    \caption{Grouped SHAP beeswarm plot averaged over classes.}
    \label{fig:shap}
\end{figure}

Finally, Fig.~\ref{fig:shap} shows the grouped SHAP summary (beeswarm) plot for the MLP-based decision model \cite{lundberg2017unified}. 
Here, features are organized into groups: the \textit{current PoI} group represents a set of variables indicating the agent’s current PoI, and the \textit{visited PoI} group represents variables indicating whether each PoI has been previously visited. 
Each point represents one sample, and the horizontal position indicates the SHAP value (i.e., the contribution to the model output), where positive and negative values respectively increase or decrease the predicted probability. Colors indicate low-to-high feature values within each group.
Across samples, the \textit{current PoI} shows the widest spread of SHAP values, indicating that it is the most influential group and can substantially increase or decrease the model output depending on the context. 
The \textit{visited PoI} group also exhibits a broad distribution on both sides of zero, suggesting heterogeneous and potentially non-monotonic effects across samples (i.e., higher group values do not consistently correspond to either positive or negative contributions). 
In contrast, \textit{total travel distance} and \textit{time of the day} show smaller magnitudes concentrated near zero, implying more moderate, context-dependent contributions. The \textit{distance between PoI} group remains tightly centered around zero, indicating a negligible effect on the model output under the present data and grouping scheme.
Although inter-PoI distance appears to be an important decision-making feature, our observations for Wakayama Castle suggest otherwise: because the distances between areas are, in practice, short enough that walking does not impose a burden, inter-PoI distance is unlikely to constitute a salient factor in visitors’ decision-making.

\color{black}\textbf{}

%% file: 06_conclusion.tex
\section{Conclusion}
\label{sec:conclusion}

We presented a platform that predicts and visualizes the effects of introducing mobility services to improve visitor circulation, based on a human-flow digital twin. 
The digital twin integrates a multi-agent simulator capable of representing destination choice as a function of agents’ 
current state and surrounding environmental information.
For agent-level decision making, we implemented an MLP and a GNN model tuned by observed decision instances.

We evaluated the proposed framework using human-flow data collected in Wakayama Castle Park in Japan. 
The MLP model reproduced the temporal evolution of area-level population distributions with mobility introduction with cosine similarities exceeding 0.7, indicating that the platform captures the overall trend of population redistribution induced by mobility introduction. 
Furthermore, the MLP reproduced the change in population distribution caused by mobility introduction with a cosine similarity of 0.664, showing its ability to reflect mobility-induced behavioral shifts.

For future work, we will first focus on improving the fidelity of the decision-making model, particularly by explicitly incorporating waiting actions into the model structure.
Furthermore, we will extend the platform’s visualization capabilities to support real-time rendering of simulation outcomes.

%% file: ref.bib
@inproceedings{ueg2023sim,
  author    = {Masashi Uegaki and Tatsuya Amano and Hirozumi Yamaguchi},
  title     = {Simulating Urban Pedestrian Flows by Fusing Wide-Area Location Data and Spot Pedestrian Counts},
  booktitle = {Proceedings of the 21st EAI International Conference on Mobile and Ubiquitous Systems: Computing, Networking and Services},
  year      = {2024},
}

@article{ROBIN200936,
title = {Specification, estimation and validation of a pedestrian walking behavior model},
journal = {Transportation Research Part B: Methodological},
volume = {43},
number = {1},
pages = {36-56},
year = {2009},
issn = {0191-2615},
author = {Th. Robin and G. Antonini and M. Bierlaire and J. Cruz},
}

@ARTICLE{10024374,
  author={Huang, Rong and Zhao, Xuan and Yuan, Yufei and Yu, Qiang and Liu, Chengqing and Daamen, Winnie},
  journal={IEEE Transactions on Intelligent Transportation Systems}, 
  title={Modeling Pedestrian Tactical and Operational Decisions Under Risk and Uncertainty: A Two-Layer Model Framework}, 
  year={2023},
  volume={24},
  number={5},
  pages={5259-5281},
}

@article{kaziyeva2023large,
  title={Large-scale agent-based simulation model of pedestrian traffic flows},
  author={Kaziyeva, Dana and Stutz, Petra and Wallentin, Gudrun and Loidl, Martin},
  journal={Computers, Environment and Urban Systems},
  volume={105},
  pages={102021},
  year={2023},
  publisher={Elsevier}
}

@article{GAO2023103369,
title = {Towards travel recommendation interpretability: Disentangling tourist decision-making process via knowledge graph},
journal = {Information Processing \& Management},
volume = {60},
number = {4},
pages = {103369},
year = {2023},
issn = {0306-4573},
author = {Jialiang Gao and Peng Peng and Feng Lu and Christophe Claramunt and Yang Xu},
}

@article{DAI2022104484,
  title={Travel inspiration in tourist decision making},
  author={Dai, Fengwei and Wang, Dan and Kirillova, Ksenia},
  journal={Tourism Management},
  volume={90},
  pages={104484},
  year={2022},
  publisher={Elsevier}
}

@article{Mcc2015tour,
author = {Mccabe, Scott and Li, Chunxiao and Chen, Zengxiang},
year = {2015},
month = {07},
pages = {},
title = {Time for a Radical Reappraisal of Tourist Decision Making? Toward a New Conceptual Model},
volume = {55},
journal = {Journal of Travel Research},
}

@inproceedings{Dan2002sumo,
  title={SUMO (Simulation of Urban MObility); An open-source traffic simulation},
  author={Daniel Krajzewicz, Georg Hertkorn, Christian Rössel and Peter Wagner},
  booktitle={Proceedings of 4th Middle East Symposium on Simulation and Modelling},
  year={2002}
}

@inproceedings{arm2015juped,
  title={JuPedSim: an open framework for simulating and analyzing the dynamics of pedestrians},
  author={Armel Ulrich Kemloh Wagoum, Mohcine Chraibi, Jun Zhang and Gregor Lammel},
  booktitle={Proceedings of the 3rd Conference of Transportation Research Group of India},
  year={2015}
}

@inproceedings{axe2008traCI,
  title={TraCI: An Interface for Coupling Road Traffic and Network Simulators},
  author={Axel Wegener and Michal Piorkowski and Maxim Raya and Horst Hellbrück and Stefan Fischer and Jean-Pierre Hubaux},
  booktitle={Proceedings of the 11th Communications and Networking Simulation Symposium},
  month={April},
  year={2008}
}

@article{zhang2021speed,
  title={A speed-based model for crowd simulation considering walking preferences},
  author={Zhang, Sainan and Zhang, Jun and Chraibi, Mohcine and Song, Weiguo},
  journal={Communications in nonlinear science and numerical simulation},
  volume={95},
  pages={105624},
  year={2021},
  publisher={Elsevier}
}

@article{WORLE2021202,
  title={Modeling intermodal travel behavior in an agent-based travel demand model},
  author={W{\"o}rle, Tim and Briem, Lars and Heilig, Michael and Kagerbauer, Martin and Vortisch, Peter},
  journal={Procedia Computer Science},
  volume={184},
  pages={202--209},
  year={2021},
  publisher={Elsevier},
}

@article{king2022virtual,
  title={A virtual experiment on pedestrian destination choice: the role of schedules, the environment and behavioural categories},
  author={King, Christopher and Bode, Nikolai WF},
  journal={Royal Society open science},
  volume={9},
  number={7},
  pages={211982},
  year={2022},
  publisher={The Royal Society}
}

@article{jin2024exploring,
  title={Exploring the Pedestrian Route Choice Behaviors by Machine Learning Models},
  author={Jin, Cheng-Jie and Luo, Yuanwei and Wu, Chenyang and Song, Yuchen and Li, Dawei},
  journal={ISPRS International Journal of Geo-Information},
  volume={13},
  number={5},
  pages={146},
  year={2024},
  publisher={MDPI}
}

@article{dang2024hypedsim,
  title={HyPedSim: A Multi-Level Crowd-Simulation Framework—Methodology, Calibration, and Validation},
  author={Dang, Huu-Tu and Gaudou, Benoit and Verstaevel, Nicolas},
  journal={Sensors},
  volume={24},
  number={5},
  pages={1639},
  year={2024},
  publisher={MDPI}
}

@article{treuille2006continuum,
  title={Continuum crowds},
  author={Treuille, Adrien and Cooper, Seth and Popovi{\'c}, Zoran},
  journal={ACM transactions on graphics (TOG)},
  volume={25},
  number={3},
  pages={1160--1168},
  year={2006},
  publisher={ACM New York, NY, USA}
}

@article{yang2024co,
  title={A co-simulation system that integrates MATSim with a set of external fleet simulations},
  author={Yang, Hai and Wong, Ethan and Davis III, Haggai and Chow, Joseph YJ},
  journal={Simulation Modelling Practice and Theory},
  volume={134},
  pages={102957},
  year={2024},
  publisher={Elsevier}
}

@article{horl2021integrating,
  title={Integrating discrete choice models with MATSim scoring},
  author={H{\"o}rl, Sebastian},
  journal={Procedia Computer Science},
  volume={184},
  pages={704--711},
  year={2021},
  publisher={Elsevier}
}

@article{li2024agent,
  title={Agent-based digital traffic model generation for regions facing data scarcity using aggregated cellphone data: a case study for Brussels},
  author={Li, Jingjun and Rombaut, Evy and Vanhaverbeke, Lieselot},
  journal={International Journal of Digital Earth},
  volume={17},
  number={1},
  pages={2407046},
  year={2024},
  publisher={Taylor \& Francis}
}

@article{horni2011high,
  title={High-resolution destination choice in agent-based demand models},
  author={Horni, Andreas and Nagel, Kai and Axhausen, Kay W},
  journal={Arbeitsberichte Verkehrs-und Raumplanung},
  volume={682},
  year={2011},
  publisher={IVT, ETH Zurich}
}

@article{helbing1991mathematical,
  title={A mathematical model for the behavior of pedestrians},
  author={Helbing, Dirk},
  journal={Behavioral science},
  volume={36},
  number={4},
  pages={298--310},
  year={1991},
  publisher={Wiley Online Library}
}

@article{BADIA2023811,
title = {Shared e-scooter micromobility: review of use patterns, perceptions and environmental impacts},
journal = {Transport Reviews},
volume = {43},
number = {5},
pages = {811-837},
year = {2023},
issn = {0144-1647},
author = {Hugo Badia and Erik Jenelius}
}

@article{yang2021tourists,
  title={Tourists on shared bikes: Can bike-sharing boost attraction demand?},
  author={Yang, Yang and Jiang, Lan and Zhang, Zili},
  journal={Tourism Management},
  volume={86},
  pages={104328},
  year={2021},
  publisher={Elsevier}
}

@inproceedings{mishra2024balancing,
  title={Balancing privacy and planning: using counterfactuals to predict and optimize tourism in Wakayama City},
  author={Mishra, Malvika and Kala, Srikant Manas and Amano, Tatsuya and Yamaguchi, Hirozumi},
  booktitle={Proceedings of the 2nd ACM SIGSPATIAL international workshop on geo-privacy and data utility for smart societies},
  pages={25--30},
  year={2024}
}

@inproceedings{naeem2026questionnaire,
author = {Naeem, Rami and Manas, Srikant and Amano, Tatsuya and Yamaguchi, Hirozumi},
title = {A Questionnaire-Only Counterfactual Machine Learning Approach to Assess the Spatial Impact of Green Mobility Vehicles in Urban Parks: A Wakayama Castle Case Study},
year = {2026},
isbn = {9798400719691},
publisher = {Association for Computing Machinery},
comment = {url = {https://doi.org/10.1145/3737611.3776620}, doi = {10.1145/3737611.3776620},},
booktitle = {Companion Proceedings of the 27th International Conference on Distributed Computing and Networking},
pages = {120–125},
numpages = {6},
location = {
},
series = {ICDCN Companion '26}
}

@article{nag2025exploring,
  title={Exploring digital twins for transport planning: a review},
  author={Nag, Dipanjan and Brandel-Tanis, Freyja and Pramestri, Zakiya Aryana and Pitera, Kelly and Fr{\o}yen, Yngve Karl},
  journal={European Transport Research Review},
  volume={17},
  number={1},
  pages={15},
  year={2025},
  publisher={Springer}
}

@article{dozza2022data,
  title={A data-driven framework for the safe integration of micro-mobility into the transport system: Comparing bicycles and e-scooters in field trials},
  author={Dozza, Marco and Violin, Alessio and Rasch, Alexander},
  journal={Journal of safety research},
  volume={81},
  pages={67--77},
  year={2022},
  publisher={Elsevier}
}

@article{lundberg2017unified,
  title={A unified approach to interpreting model predictions},
  author={Lundberg, Scott M and Lee, Su-In},
  journal={Advances in neural information processing systems},
  volume={30},
  year={2017}
}
